\documentclass[lettersize,journal]{IEEEtran} 
\IEEEoverridecommandlockouts
\usepackage{cite}
\usepackage{amsmath,amssymb,amsfonts}
\usepackage{algorithmic}
\usepackage[linesnumbered, ruled, vlined]{algorithm2e}
\usepackage{multirow} 
\usepackage{graphicx}
\usepackage{caption}
\usepackage{subcaption}
\usepackage{mwe}
\usepackage{textcomp}
\usepackage{xcolor}
\usepackage{booktabs}
 \usepackage{tabularx}
 \usepackage{siunitx}
 \usepackage{multirow}
 \usepackage{comment}
 \usepackage{tikz}
\usepackage{booktabs}
\usepackage{array}
\usepackage{url}
 \usepackage{float}
\usepackage[ruled,vlined,linesnumbered]{algorithm2e}
\graphicspath{ {D:/Shankar Ghosh/Desktop/images/} }
\def\BibTeX{{\rm B\kern-.05em{\sc i\kern-.025em b}\kern-.08em
    T\kern-.1667em\lower.7ex\hbox{E}\kern-.125emX}}
\newtheorem{definition}{Definition}
\begin{document}

\title{RL+AHP: A Novel Reinforcement Learning driven AHP for Slice Aware mode selection in D2D enabled Heterogeneous Networks}

\author{Souvik Deb, Sumita Majhi, Shankar K. Ghosh, {\it Member, IEEE}, Avirup Das, {\it Member, IEEE}, Rajib Mall, Sridevi S and Jacob Augustine
  \thanks{Souvik Deb is affiliated with School of Computer Science, University of Petroleum and Energy Studies, Dehradun, India. Email: souvik.deb@upes.ac.in}
  \thanks{Sumita Majhi is affiliated with Department of Computer Science and Engineering, Indian Institute of Technology Guwahati, Guwahati, India. Email: \{sumit176101013\}@iitg.ac.in}
  \thanks{Shankar K. Ghosh and Rajib Mall are affiliated with Department of Computer Science and Engineering, Shiv Nadar Institution of Eminence, Delhi NCR, India. Emails: \{shankar.it46, rajib.mall\}@snu.edu.in.}
 \thanks{Avirup Das is affiliated with Singapore University of Technology and Design, Singapore. Emails: avirup1987@gmail.com.}
 \thanks{Sridevi S and Jacob Augustine are affiliated with School of Computer Science and Engineering, Presidency University, Bengaluru, India. Emails: \{sridevi.svs2809, jacob.ku.augustine\}@gmail.com.}
 \thanks{Shankar K. Ghosh and Avirup Das will be acting as corresponding authors.
 
 A preliminary version of the manuscript has been archived on 15th March 2026 (DOI: \protect\url{https://arxiv.org/pdf/2603.14551}).}
}
\maketitle

\begin{abstract}
The mode selection problem in device-to-device communication (D2D) enabled Fifth generation (5G) heterogeneous networks (HetNet) aims prioritizing four key performance indicators (KPIs) namely data rate, latency, reliability and jitter across three slices: enhanced mobile broadband (eMBB), ultra reliable low latency (uRLLc) and massive machine type communications (mMTC). Such priority assignment must be \emph{traded off} among three access technologies, i.e., Long Term Evolution advanced (LTE-A), New Radio (NR) and D2D, while minimizing handover frequency. In existing mode selection approaches for HetNet, slice specific quality of service (QoS) requirements are largely ignored. In this work, a novel mode selection algorithm is proposed by combining a two level Analytic Hierarchy Process (AHP) with a Reinforcement Learning (RL) method. While the two level AHP facilitates decision making based on multiple criteria (i.e., KPIs) and options (i.e., LTE-A, NR, D2D mode), the RL approach computes the weights of each criteria based on the feedback from the environment. Simulation results show that our proposed algorithm outperforms related works in terms of the major KPIs for all three slices. For eMBB applications, our approach increases throughput by $33\%$; for uRLLc applications, our approach significantly decreases latency and BER ($27\%$ and $10\%$ respectively) and for mMTc applications, our approach significantly decreases latency ($44\%$). Moreover, it has been shown that the proposed RL+AHP approach outperforms the existing DRL based approaches in terms of CPU usage when the number of criteria is reasonably low ($<6$). 
\end{abstract}

\begin{IEEEkeywords}
5G, Device-to-device communication, mode selection, slice awareness, Reinforcement Learning, Analytic Hierarchy process.
\end{IEEEkeywords}

\section{Introduction}
\begin{figure}
    \centering
    \includegraphics[width=0.75\linewidth]{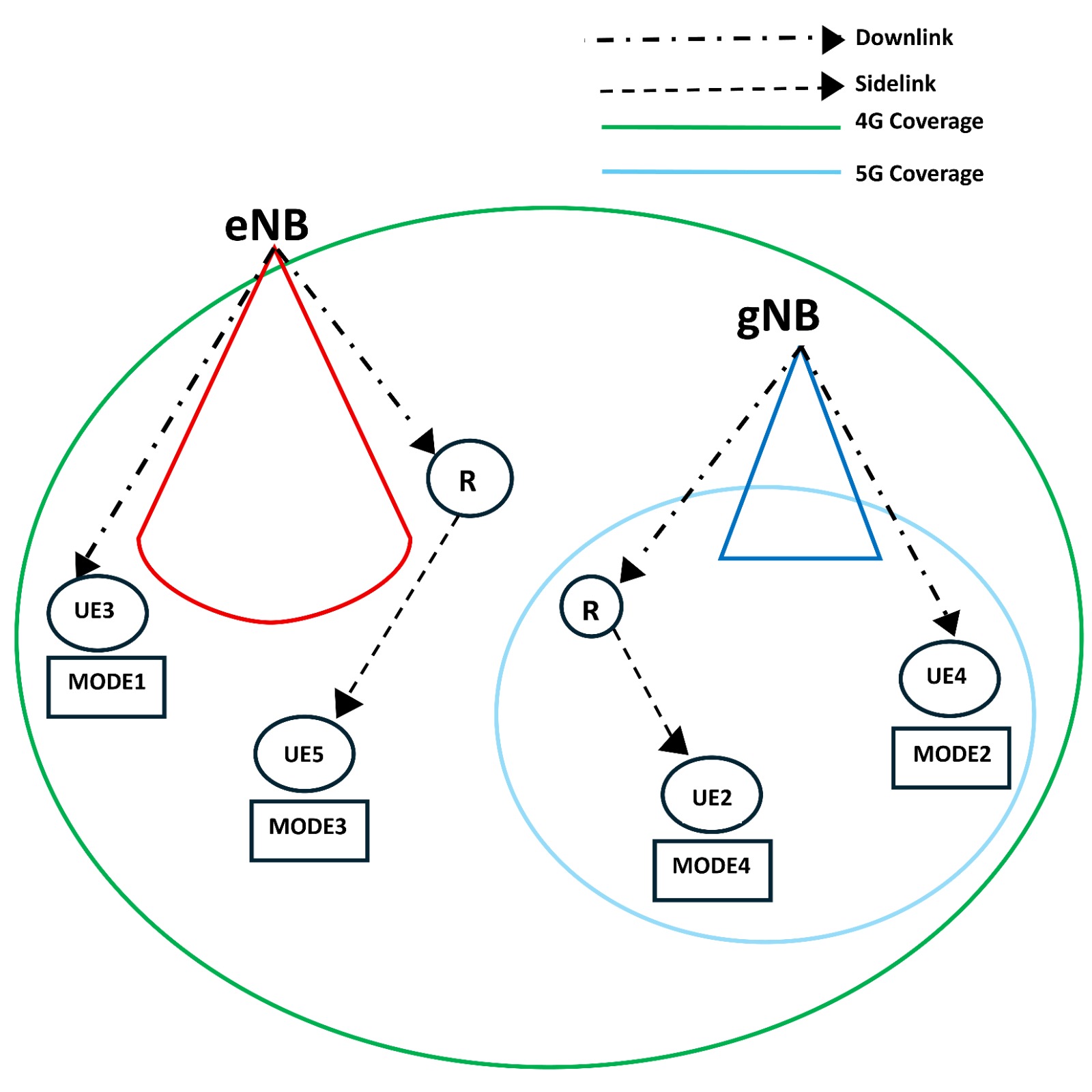}
    \caption{An example of D2D enabled HetNet:  UE 3 is communicating in Mode-1;  UE 4 is communicating in Mode-2; UE 5 is communicating in Mode-3; and UE 2 is communicating using Mode-4.}
    \label{D2D5G}
\end{figure}

In Non-standalone (NSA) deployment of fifth generation (5G) cellular network, Long Term Evolution advanced (LTE-A) and New Radio (NR) systems co-exist \cite{NSA}. Therein, LTE-A macro-cells provide ubiquitous coverage, whereas ultra dense deployment \cite{udn2021} of New Radio (NR) small cells provide additional capacity in high-traffic areas. While roaming across such a heterogeneous network (HetNet) consisting of both LTE-A and NR systems, user equipments (UEs) may have quality of service (QoS) requirements specific to any one of the following slices: enhanced mobile broadband (eMBB), ultra reliable low latency communication (uRLLc) and massive machine type communications (mMTC) \cite{emBBuRLLc}. The eMBB applications require high bandwidth resources to achieve transmission rates upto $20$ Gbps; the uRLLc applications require response time less than $1$ ms with reliability greater than $99.99\%$; and the mMTC services are usually up-link centric requiring low data rate and jitter. In such HetNet scenario, key challenges in satisfying the service requirements of the UEs are: reduced Reference Signal Received Power (RSRP) at the cell edges, channel fading, user mobility, blockage of millimeter wave (mmWave) links by dynamic obstacles and inter-cell interference caused by dense deployment of NR small cells. To ensure service demands in such scenarios, device-to-device (D2D) communication \cite{D2D6G} was proposed, where intermediate relays are used to enhance channel quality. The D2D technology is known to be effective in satisfying the quality of services (QoS) for all three slices in 5G cellular network \cite{erricson}, \cite{D2D6G2020}, \cite{IEEETVT2019}. In this context, coupling of D2D pairs with appropriate radio access networks (e.g., LTE, NR) is of paramount importance.

A D2D enabled HetNet is shown in Fig. \ref{D2D5G}. Therein, a UE can communicate in four \emph{modes}: {\bf (a) direct communication with LTE-A macro-cell (namely Mode 1); (b) direct communication with NR small cell (namely Mode 2);  (c) Communication with LTE-A macro-cell via an intermediate relay (namely Mode 3); (d) Communication with NR small cell via an intermediate relay (namely Mode 4);} Mode 1 and Mode 2 are often referred to as cellular modes; whereas Mode 3 and Mode 4 are referred to as dedicated (or reuse) modes depending on the usage of resources \cite{1}. In Fig. \ref{D2D5G}, UE 3 is communicating in Mode 1;  UE 4 is communicating in Mode 2; UE 5 is communicating in Mode 3; and UE 2 is communicating using Mode 4. To enhance the QoS for these UEs, efficient mode selection is extremely important. 

The mode selection mechanism involves several {\it opposing factors}. For example, the LTE-A system provides lower data rate than NR (because of smaller bandwidth) but offers higher reliability than the NR systems. This is because the blockage of mmWave links by dynamic obstacles lead to high packet losses, thus increasing number of retransmissions. In LTE-A system, latency is higher than NR systems due to higher propagation and queuing delays. In D2D communication (i.e., Mode 3 and Mode 4), data rate is high at the cell edges as compared to that of direct LTE and NR communications, but reliability may be less due to the blockage of D2D links by dynamic obstacles. Switching from one mode to another (e.g., Mode 1 $\leftrightarrow$ Mode 2, Mode 1 $\leftrightarrow$ Mode 4) may cause handovers between access points, leading to high signaling overhead. In such a prevailing situation, a research question we target to address is: {\it Given a target QoS, which of the aforementioned modes are preferred in a D2D  enabled HetNet scenario?}    

\subsection{Discussions on prior work}
Past research on mode selection offers solutions from various angles, but with notable shortcomings as discussed below. 

\paragraph{Fixed threshold based approaches}  A large body of related research on mode selection is based on fixed threshold assumptions. The joint mode selection, relay selection, and resource allocation (JMSRA) algorithm proposed in \cite{JMSRA} is based on a QoS model which imposes a static limit on data rate but does not consider the key KPIs for distinguishing 5G slices. Multi-attribute decision-making (MADM) methods have extensively been used for network selection in HetNet \cite{b2, MCDMHetNet1, MCDMHetNet2, MCDMHetNet3, MADM}. These algorithms are, however, configured for 4G HetNets comprising of Universal Mobile Telecommunications System (UMTS), Global System for Mobile Communications (GSM), LTE-A and Wireless Local Area Network (WLAN); and do not account 5G specific requirements, i.e., mmWave bands (24 GHz-52 GHz) with blockage sensitivity, impact of dynamic obstacles, D2D communication capabilities, frequent handovers and slice specific KPI requirements. These factors affect RSRP statistics which is crucial for any decision making. In \cite{drissi2017multi}, a multi‑criteria real‑time network selection framework has been proposed for 5G HetNet using Fuzzy Analytic Hierarchy Process (FAHP). Therein, static weights have been used to characterize application types (conversational, streaming, interactive, background). Moreover, D2D communication is not considered. Meanwhile \cite{roy2025fuzzy} proposes a FAHP based network selection algorithm for 5G HetNet (consisting of 2G, 3G, 4G, 5G and WiFi) to serve voice, video, and image transmission, using a Dynamic Pairwise Comparison Matrix (D‑PCM) where exponential, sigmoid and step functions are used to capture service specific QoS requirements. In \cite{sridevi2022analytic}, the authors propose an AHP based UE-UAV association method that balances power, secrecy, and packet loss trade-offs between UAVs at different altitudes in beyond 5G networks. However, the priority weights generated by the AHP based method do not adapt to dynamic network conditions. In these existing works  \cite{b2, MCDMHetNet1, MCDMHetNet2, MCDMHetNet3, MADM, drissi2017multi, roy2025fuzzy}, D2D communication has not been considered.

The existing relay selection and handover methods \cite{MARelaySelection2019, modeswitching2014, QosawareD2D,  b17} also lack QoS awareness. In \cite{MARelaySelection2019}, the relay selection algorithm incorporates mobility-awareness via stochastic integer programming, but the QoS requirements and variety of radio access networks in 5G HetNet are not taken into account. The resource allocation strategies in \cite{modeswitching2014, modeselection2019, QosawareD2D} optimize spectral efficiency and delay tolerance, without explicitly accounting for slice-aware KPIs.  The handover mechanism  for D2D-enabled 5G HetNet in \cite{b17} ignores the QoS requirements, as well as Layer 3 signaling overhead. 
These fixed threshold based approaches do not account for the non-stationarity caused by UE mobility, varying channel dynamics and diverse QoS demands of different slices. In \cite{ahp_arxiv}, the mode selection problem was addressed by using a static AHP framework. However, the non-stationarity of the wireless network caused by UE mobility, channel fading and varying QoS demands over time require a framework where the AHP weights can be dynamically adjusted based on the changing network conditions. 
\paragraph{Data driven approaches} The traditional optimization techniques often struggle with the dynamics of large-scale and complex wireless networks \cite{RLAHP5}. Therefore, Reinforcement Learning (RL) is being extensively applied to real world problems. To address the drawbacks of fixed threshold based approaches, recent studies increasingly use RL to select modes intelligently in dynamic network systems \cite{1, 2, 3, 7, 8}. However, the key deficiencies in these works are the lack of slice specific QoS awareness and assumption of static topology. The Deep Reinforcement Learning (DRL) based mode selection approaches for Vehicle-to-Everything (V2X) scenarios \cite{1, 2, 3} mainly focus on power adaptation and resource allocation in different fading conditions to optimize packet reception time; however they are designed for static topology. 
DRL approaches are computationally intensive. Due to the high convergence time and resource requirements, DRL approaches are often unsuitable for constrained devices and delay sensitive applications. 

The multi-agent RL based mode selection algorithm for 5G HetNet in \cite{8} aims to optimize throughput for virtual reality (VR) services. The semantic communication approach in \cite{7} assumes reuse mode, where D2D links share the same resources causing co-channel interference. In \cite{8, 7}, slice specific QoS requirements have not been considered.

\paragraph{Hybrid RL+AHP approaches}
In the early learning stage, RL performs trial-and-error search, often leading to high convergence time and delayed reward. This is unacceptable for QoS constrained applications. Integrating Analytic Hierarchy Process (AHP) with RL increases the chance of selecting promising actions over random ones in early learning stages. The RL+AHP combination adapts to fast-changing network conditions (e.g., user velocity, traffic load variation, RSRP) based on the feedback from the environment; and therefore, performs better than static, rule-based approaches. In recent past, such integrated models have been used for handover management \cite{RLAHP1, RLAHP2}, adaptive routing \cite{RLAHP4}, throughput optimization \cite{ma2021intelligent}, network selection in HetNet \cite{RLAHP3} and real time slice control for autonomous mobility services in 5G network \cite{RL-Loop}. However, application of RL+AHP model in the context of mode selection is unprecedented in literature.

From the above discussions, it is evident that developing an unified mode selection framework for 5G HetNet accounting for diverse QoS demands of different slices, varying channel dynamics and handover costs is an important topic for further research.

\subsection{Our contributions}
The mode selection problem involves prioritizing four key performance indicators (KPIs), i.e., data rate, latency, reliability, and jitter across three 5G slices, i.e., eMBB, uRLLC and mMTC. Such priority assignment must be traded off among three access technologies, i.e., LTE-A, NR and D2D. Hybrid RL+AHP is particularly helpful in handling multi-objective decision-making under uncertainty in wireless networks \cite{RLAHP6}. Therein, the RL part excels at dynamic and trial-and-error learning of optimal policies whereas AHP systematically weigh and prioritize among conflicting criteria. In this work, our \emph{objective} is to develop a slice aware mode selection algorithm for D2D enabled 5G HetNet. Our main {\it contributions} are summarized as follows.

\begin{itemize}
    \item {\bf Problem formulation:} The mode selection problem in D2D enabled 5G HetNet is formulated as a Markov Decision Process (MDP) accounting slice specific QoS requirements, UE velocity, current mode of communication and Signal-to-Interference-plus-Noise Ratio (SINR) at the UE. Therein, actions represent mode selections and reward represents the extent upto which the QoS is met. Our goal is to determine the policy (i.e., modes to be selected over time) to optimize long term discounted reward (i.e., QoS achieved). 
    \item {\bf Novel RL+AHP based mode selection:} To solve the mode selection problem, the AHP technique has been combined with an RL approach namely Combinatorial Adversarial Multi-armed bandit (CAMAB) approach. The CAMAB part is responsible for finding the optimal combination of entries in the Level 0 Pairwise Comparison Matrix (PCM) of the AHP. While the proposed two-level AHP approach facilitates decision making based on multiple criterion and options (shown in Fig. S1), the CAMAB approach determines the weights of each criteria based on the feedback from the environment. In \textbf{Level 0} of the proposed AHP, the QoS requirements of application classes have been characterized using four key performance indicators (KPIs) namely \emph{data rate}, \emph{latency}, \emph{reliability} and \emph{jitter} (criteria for the AHP). The \textbf{Level 1} of the proposed AHP framework performs pairwise comparisons between the modes (options for the AHP). Based on the pairwise comparisons in Level 0 and Level 1, the proposed AHP computes a priority vector for a given application class. Next, the priority vector (computed by the RL+AHP method) is combined with the RSRPs from LTE-A, NR and  relay devices, to incorporate the impact of UE mobility. Thereafter, mode selection is performed based on the computed priorities. Finally, feedback from the environment is collected in terms of the perceived QoS, which is further used by the CAMAB algorithm to compute the criteria weights of the AHP (at Level 0). 
  
    \item {\bf Performance studies:} The performance of our proposed RL+AHP mode selection approach has been compared with three existing mode selection approaches: a SDN based \cite{b7}, a RSRP based \cite{RSRP_baseline} and the JMSRA \cite{JMSRA}. The simulation results reveal that our proposed algorithm significantly outperforms the existing approaches in terms of the major KPIs for all three slices, while ensuring appropriate trade-offs among the opposing factors by choosing modes appropriately. Furthermore, it has been shown that the proposed RL+AHP approach outperforms the existing DRL based approaches in terms of CPU usage when the number of criteria is reasonably low ($<6$). Beyond that limit, the RL+AHP and DRL based approaches show similar performances.
\end{itemize}
For eMBB applications, our approach increases throughput by $33\%$; for uRLLc applications, our approach significantly decreases latency and BER ($27\%$ and $10\%$ respectively); for mMTc applications, and significantly decreases latency ($44\%$), when compared to the related works. The proposed RL+AHP approach reduces the CPU usage time by $85\%$ and $75\%$ when the number of criteria are $4$ and $5$ respectively.  

Rest of this paper is organized as follows. In section \ref{sysmodel}, system model has been presented. In Section \ref{probform}, the mode selection problem has been formulated as MDP. In section \ref{pho}, the proposed RL+AHP approach for mode selection approach has been described in detail. Based on the priority computed by AHP, a mode selection algorithm has been proposed in section \ref{pho}. Section \ref{results} reports the results of the performance studies carried out on our algorithm based on extensive system level simulations. Finally, section \ref{con} concludes the work.

\section{System Model} \label{sysmodel}
The considered 5G HetNet consists of $N$ number of evolved Node Bs (eNBs) and $M$ number of Next Generation Node B (gNBs) as shown in Fig. \ref{D2D5G}. Without loss of generality, the set of Base Stations (BSs) is denoted by $\mathcal{S} = \{1, 2, \dots, N, N+1, \dots, N+M\}$, where the indices $\{1, \dots, N\}$ represent eNBs providing ubiquitous coverage, and the indices $\{N+1, \dots, N+M\}$ represent gNBs\footnote{Subsequently, "eNBs" and "gNBs" will be interchangeably used with "macro cell" and "small cells" respectively.}, providing high data rates in hot-spot areas. The set of UEs is represented by $\bar{D} = \{1, 2, \dots, D\}$. These UEs are capable of both cellular and D2D communications. We assume that D2D sidelinks are using mmWave frequency bands and reuse the downlink resources of the BSs to enhance spectral efficiency as considered in \cite{JMSRA, 1, MARelaySelection2019}. Relay selection is done considering UE mobility as done in \cite{MARelaySelection2019}.  

\subsection{Modeling SINR}
In the considered scenario, a UE can be served in four distinct modes namely \textbf{Mode-1}, i.e., direct communication with LTE-A macro cell; \textbf{Mode-2}, i.e., direct communication with NR small cell; \textbf{Mode-3}, i.e., communication with LTE-A via an intermediate D2D relay; and \textbf{Mode-4}, i.e., communication with NR via an intermediate D2D relay. The signal-to-interference-plus-noise ratio (SINR) for these modes are modeled below.

\begin{table}[htbp]
\centering
\caption{Symbols and Definitions}
\label{tab:symbols}
\begin{tabular}{|c|p{6.3cm}|}
\hline
\textbf{Symbols} & \textbf{Definitions} \\
\hline
$I_{i,l}^B(t)$ & Interference received at $D_i$ while \\ 
    & connected with BS $l$ at time $t$. \\
\hline
$S$ & Set of BSs. \\
\hline
$\overline{D}$ & Set of UEs. \\
\hline
$P_{l,i}^{T}(t)$ & Transmitting power from BS $l$ to  UE $i$ at time $t$. \\
\hline
$P_{j,k}(t)$ & Transmitting power from  UE $j$ to UE $D_k$ at time $t$. \\
\hline
$G_{l,i}^{T}(t)$ & Channel gain of BS $l$ to UE $i$ downlink at time $t$. \\\hline
$P^R_{l,i}(t)$ & Received power at UE $i$ from BS $l$ at time $t$.\\
\hline
$G_{j,k}(t)$ & Channel gain of UE $j$ to UE $k$ link at time $t$ \\
\hline
$\sigma^{2}$ & Thermal noise power. \\
\hline
$f_{j,k}^{l}(t)$ & Binary indicator representing if $(j,k)$ pair \\ 
   & reuses downlink resources of BS $l$. \\
\hline
$p_{l,i}^{LOS}(t)$ & Probability of LOS from BS $l$ to UE $i$ at time $t$.\\\hline
$\gamma_{i,l}(t)$ & SINR at UE $i$ when receiving from BS $l$. \\
\hline
$\gamma_{j,k}^{l}(t)$ & SINR at UE $j$ when receiving from UE $k$. \\
\hline
$R_{ij}(t)$ & Achievable data rate on link from transmitter $i$ \\
& to receiver $j$ at time $t$ \\
\hline
$N_{\text{RB}}$ & Number of allocated resource blocks (RBs) \\
\hline
$N_{\text{SC}}$ & Number of subcarriers per resource block. \\
\hline
$N_{\text{sym}}$ & Number of symbols per resource block. \\
\hline
$\text{CR}_{i,j}$ & Coding rate used on link UE $i$ to UE $j$. \\
\hline
$M_{i,j}$ & Modulation order (number of states) on link UE $i$ to UE $j$. \\
\hline
$T_{\text{RB}}$ & Duration of one resource block (symbol time) \\
\hline
PCM$_{ij}$ & Entry in $i^{th}$ row and $j^{th}$ column of PCM \\
\hline
$\eta$ & Learning rate for CAMAB \\
\hline
T & Total number of time steps \\
\hline
$\Delta$ & Time discretizations granularity \\
\hline
\end{tabular}
\end{table}

\paragraph{SINR computation for cellular downlink (Mode 1 and Mode 2)}

The interference $I_{i,l}^B(t)$ received at UE $i$ while connected with BS $l$ at time $t$ is computed as $\displaystyle \sum_{a \in S \setminus \{l\}} P_{a,i}^T(t) \times G_{a,i}^T(t) + \sum_{j,k \in \bar{D} \setminus \{i\}} P_{j,k}(t) \times G_{j,i}(t) \times f_{j,k}^l(t)$, where $P_{l,i}^T(t)$ denotes transmitting power from BS $l$ to UE $i$, $P_{j,k}(t)$ denotes transmitting power from UE $j$ to UE $k$, $G_{l,i}^T(t)$ denotes the channel gain between BS $l$ and UE $i$, $G_{j,i}(t)$ denotes channel gain of the UE $j$ to UE $i$ link ($i \in \bar{D}$ and $l \in \{1, 2, \dots, N+M\}$). Here $\sum_{a \in S \setminus \{l\}} P_{a,i}^T(t) \times G_{a,i}^T(t)$ represents the interferences from other base stations and $\sum_{j,k \in \bar{D} \setminus \{i\}} P_{j,k}(t) \times G_{j,i}(t) \times f_{j,k}^l(t)$ represents interference from sidelinks sharing the same frequency with BS $l$ to UE $i$ downlink. The interference coordination function $f_{j,k}^l(t)$ is defined as:
\begin{small}
    \begin{equation*}
f_{j,k}^l(t) = 
\begin{cases} 
1 & \text{if  UE j to UE k link reuses BS $l$ to UE i downlink}  \\
0 & \text{otherwise}
\end{cases}
\end{equation*}
\end{small}

\textbf{Mode-2} relies on mmWave communication. Due to high pathloss in mmWave, received power at UE $i$ become negligible when a dynamic obstacle blocks the Line of Sight (LOS) between gNB $l$ and UE $i$. Now, in the presence of dynamic obstacles, $p_{l,i}^{LOS}(t)$ the probability that LOS exists between gNB $l$ and UE $i$ at time $t$, can be computed following 3GPP/ITU standard as follows \cite{mmWaveRLF}:

\begin{equation}
    p_{l,i}^{LOS}(t)=\min(20/d_{li}(t),1)(1-\epsilon^{\frac{-d_{li}(t)}{39}}) + \epsilon^{\frac{-d_{li}(t)}{39}}.
    \label{losprob}
\end{equation}
\noindent Here, $d_{li}(t)$ is the distance (in meters) between gNB $l$ and UE $i$ at time $t$. Therefore, the received power $P^R_{l,i}(t)$ at UE $i$ from gNB $l$ is computed as:

\begin{equation}
    P^R_{l,i}(t)=\begin{cases}
        P_{l,i}^T(t) \times G_{l,i}^T(t)\times  p_{l,i}^{LOS}(t) & \text{for gNBs}\\
         P_{l,i}^T(t) \times G_{l,i}^T(t) & \text{for eNBs}
    \end{cases}
\end{equation}

Accordingly, the SINR $\gamma_{i,l}^l(t)$ received at UE $i$ from gNB $l$ at time $t$ is computed as:
     \begin{equation}
      \gamma_{i,l}^l(t) = 
    \frac{P^R_{l,i}(t)}{I_{i,l}^B(t) + \sigma^2},
\label{eq:downlink_sinr}      
     \end{equation}

\noindent where $\sigma^2$ represents the thermal noise power.

\subsection{SINR computation for relay-assisted side links (Mode 3 and Mode 4)}
In the relay-assisted modes, the interference $I_{j,k}(t)$ received at  UE $k$ from the relay UE $j$ is computed as: $\sum_{a \in S \setminus \{l\}} P_{a,k}^{T}(t) \times G_{a,k}^{T}(t) + \sum_{j',k' \in \bar{D} \setminus \{j,k\}} P_{j',k'}(t) \times G_{j',j}(t)$. Here  $P_{j,k}(t)$ represents the transmitting power from the relay UE $j$ to UE $k$ and  $G_{j,k}(t)$ denotes the channel gain of the sidelink between UE $j$ and UE $k$. The former term $\sum_{a \in S \setminus \{l\}} P_{a,k}^{T}(t) \times G_{a,k}^{T}(t)$ represents interference from all base stations and the latter term $\sum_{j',k' \in \bar{D} \setminus \{j,k\}} P_{j',k'}(t) \times G_{j',j}(t)$ represents the interference caused by other active D2D pairs in the network. Accordingly, the SINR $\gamma_{j,k}(t)$ at UE $k$ while connected with the relay UE $j$ is computed as: 

\begin{equation}
\gamma_{j,k}(t) = \frac{P_{j,k}(t) \times G_{j,k}(t)\times p_{j,k}^{LOS}(t)}{I_{i,k}(t) + \sigma^2}.
\label{eq:sidelink_sinr}
\end{equation} 
\noindent Here, $p_{j,k}^{LOS}(t)$ is the probability of LoS between UE $j$ and UE $k$ at time $t$ as computed in \eqref{losprob}. Subsequently, based on \eqref{eq:downlink_sinr} and \eqref{eq:sidelink_sinr}, data rates for different modes are modeled as follows.

\subsection{Modeling of Data Rates}
The mapping between the SINR and the Modulation and Coding Scheme (MCS) has been done according to 3GPP \cite{JMSRA}. In general, the data rate $R_{ij}(t)$ between transmitter $i$ and receiver $j$ at time $t$ computed as:
\begin{equation}
R_{ij}(t) = N_{RB}(t) \frac{N_{SC} N_{sym} CR_{i,j} \log_{2} M_{i,j}}{T_{RB}},
\end{equation}
where $N_{RB}(t)$ is the number of allocated RBs at time $t$, $CR_{i,j}$ is the coding rate which depends on the SINR as computed in Equations \eqref{eq:downlink_sinr} and \eqref{eq:sidelink_sinr}, and $M_{i,j}$ is the number of modulation states. Here the possible transmitters are eNBs, gNBs and UEs; on the other hand possible receivers are only UEs as we focus only on downlink. The RB allocation is done using standard proportional fair access mechanism \cite{packetscheduling2013}. Based on this system model, in the next section, we formulate the mode selection problem as MDP.

\section{MDP formulation}\label{probform}
In D2D enabled 5G HetNets, efficient mode selection is essential to ensure that UEs satisfy the QoS requirements of each slice, i.e., eMBB, uRLLc and mMTc. Here time is discretized with a granularity $\Delta$. We assume that mode selection occurs at the beginning of a time step. Switching from one mode to another is a sequential decision making problem under the uncertainty of stochastic channel conditions, positions of the dynamic obstacles and UE mobility. Therefore, for a given slice, the mode selection problem is formulated as an MDP whose optimal policy maximizes the long term QoS. Here, for a particular application, QoS is defined as a $4$-tuple $\left\lbrace Th_d, Th_r, Th_l, Th_j\right\rbrace$ representing thresholds on KPIs; $Th_d$ represents the lower bounds on data rate,  $Th_r$ represents the lower bound reliability, $Th_l$ represents the upper bound on latency and $Th_j$ represents the upper bound on jitter. 
\paragraph{State Space} At each time step $t$, a UE decides to connect to a particular mode based on the observation about its current mode of communication, current quality of experience (i.e., whether or not the required KPIs are satisfied), velocity and the current channel conditions. Therefore, a state in the state space must consist of all the aforementioned observations. Formally, the state at time step $t$ is represented by the following vector:
\begin{equation*}
   s(t)=<Q(t),\;v(t),\; m(t),\; \beta(t)>.
\end{equation*}
Here, $Q(t)$ represents the indicator of whether the UE has achieved the desired quality of experience (i.e., the thresholds on data rate, reliability, latency or jitter are satisfied) at the current time $t$. These thresholds vary depending on the slices. Here, $Q(t)=1$ if the desired QoS is achieved, and $0$ otherwise; $v(t)$ is the velocity of the UE at time step $t$; $m(t)$ represents the current mode of communication and $\beta(t)$ is the SINR at the UE at time $t$. It may be noted that $\beta(t)$ and $v(t)$ are \emph{continuous}, whereas $Q(t)$ and $m(t) \in \left\lbrace 1, 2, 3, 4 \right\rbrace$ are \emph{discrete}. Depending on $m(t)$, the value of $\beta(t)$ is computed using Equations \eqref{eq:downlink_sinr} and \eqref{eq:sidelink_sinr}. The resulting state space is continuous.

\paragraph{Action space} The action space $Y=\{y_1, y_2, y_3, y_4\}$ constitutes of $4$ actions, where action $y_i$ corresponds to selection of mode $i$.

\paragraph{Reward} After choosing action $c_i$ at time $t$, a reward $R(t+1)$ is received at time $t+1$. The reward $R(t+1)$ represents the \emph{extent} upto which the QoS is achieved and the overhead caused by handovers.

\paragraph{Policy}
Based on the state $s(t)$, the policy function $\pi: S \rightarrow A$ decides the action to be taken. The goal of the policy function is to maximize the expected long term discounted reward, i.e., $\mathbb{E} \left\lbrace \displaystyle \sum_{i=0}^{\infty} \gamma^i \times R(t+i)\right\rbrace$, where $\gamma$ represents the discount factor. 

\paragraph{Transition probabilities} After executing an action and observing the corresponding reward $R(t)$  at time step $t$, the state of the environment transitions to $s(t+1)$ at time step $t+1$. The transition probability $\mathbb{P}\left\lbrace s(t+1)|s(t), A(t) \right\rbrace$ from state $s(t)$ to state $s(t+1)$ depends on a number of stochastic factors such as channel condition, UE velocity and UE position at time $t+1$, which are {\it unknown} in advance, rendering the MDP to be model free. 

\subsection{Solution approach}
Given the model free nature of the MDP and the continuous state space, Deep Reinforcement Learning (DRL) is a good candidate for the learning of optimal policy. However, DRL approaches are computationally intensive. Due to the high convergence time and resource requirements, DRL approaches are often unsuitable for constrained devices and delay sensitive applications. {\bf Therefore, in this work, we have combined a Multi-criteria decision making method namely Analytic Hierarchy Process (AHP) with a suitable RL approach for faster convergence.} Integrating AHP with RL increases the chance of selecting promising actions over random ones in early learning stages. The AHP breaks the complex structure of the problem into a hierarchy of goals, criteria and alternatives. Therein, the KPIs (data rate, reliability, latency and jitter) are considered as criteria; whereas the modes are considered as options. The AHP technique also provides a built in consistency validation mechanism through the consistency ratio check, which is absent in DRL based methods resulting in more exploration.

A key deficiency of AHP technique is its static nature. That is the final ranking is made based on the inputs to the PCM given at the beginning of the process. Such ranking schemes are completely unaware of the environmental changes, leading to the ranking anomaly problem. To deal with this drawback, the AHP has been combined with a RL approach namely CAMAB. The CAMAB part excels at dynamic and trial-and-error learning of optimal policies. In the next section, we present our proposed mode selection approach.

\begin{figure*}
    \centering
    \includegraphics[width=12cm, height=14cm]{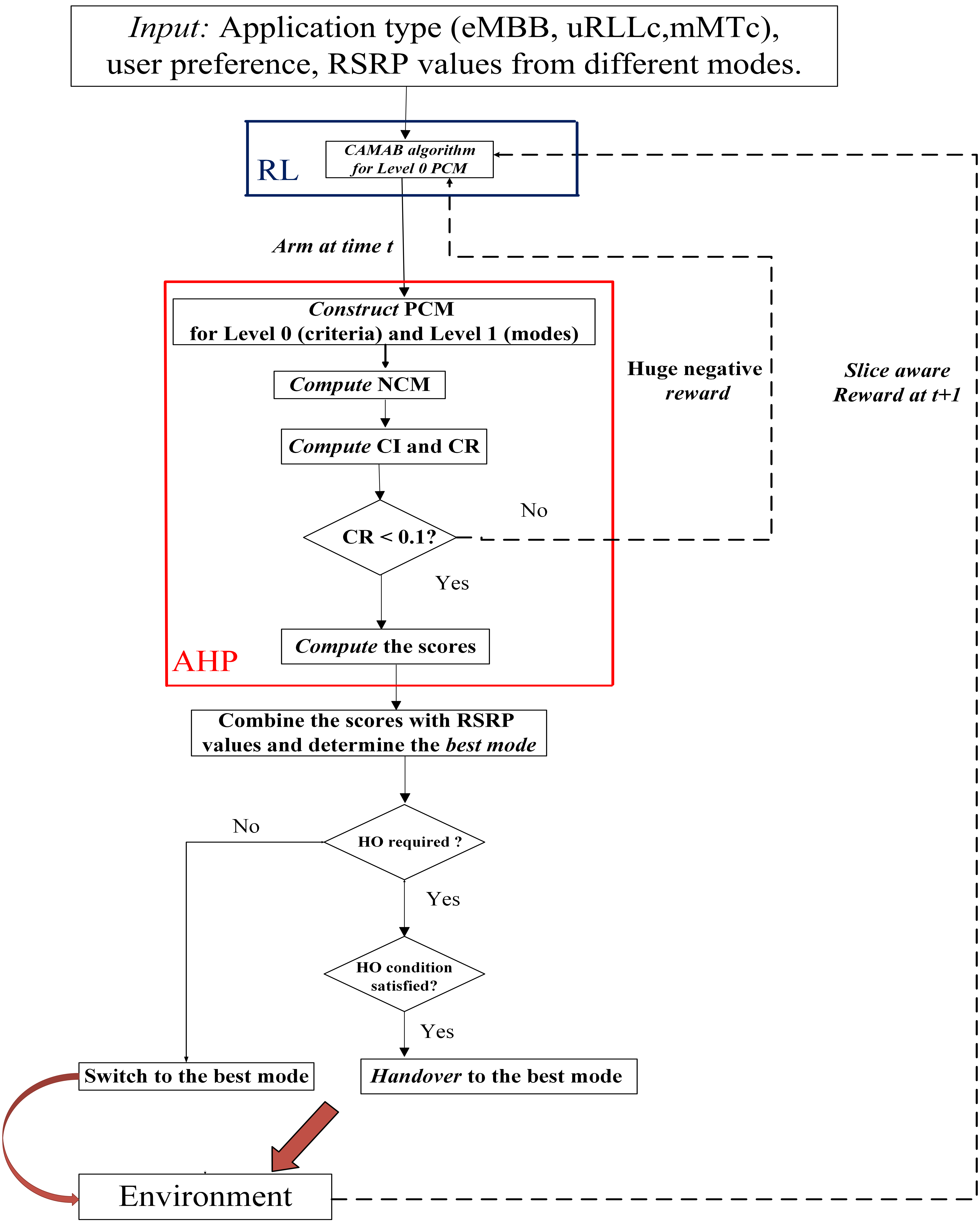}
    \caption{Flowchart representation of the proposed mode selection approach. Here, the wide arrows represent the actions on the environment and the dotted arrows represent the feedback from the environment.}
    \label{Flowchart}
\end{figure*}

\section{Proposed RL+AHP method for mode selection} \label{pho}
The proposed method for mode selection has two key components: (a) two-level AHP to prioritize between different modes, i.e., LTE-A only (Mode 1), NR only (Mode 2) and D2D communications (Mode 3 and Mode 4), based on the QoS requirements of different application classes, i.e., eMBB, uRLLc and mMTC; and (b) a RL approach namely CAMAB which is responsible for finding the optimal combination of entries in the Level 0 PCM of the two-level AHP. The operation of the proposed method is shown in Flowchart format in Fig. \ref{Flowchart}. Therein, the dotted arrows represent the feedback from the environment and the wide arrows represent the actions on the environment. 

In the proposed AHP technique, the QoS requirements for different application classes have been represented using pairwise comparison matrices consisting of four criteria namely \emph{data rate}, \emph{latency}, \emph{reliability} and \emph{jitter} at {\bf Level $0$}. The  priority values has been shown in Table S1. These values represent the priority of a criteria in a scale of $10$, where the priority is directly proportional to the values. These weight assignments to different criteria are done based on the requirement of the application as well as user preferences \cite{b2} (shown in Table \ref{cr_weights}). Next in {\bf Level $1$}, pairwise comparisons are done between the modes for each criteria. Finally, the ranks are computed by combining the criteria weights from Level $0$ and Level $1$.

On the other hand, the CAMAB's arms serve as the entries of the PCM matrix of the AHP and the reward reflects the impact of the input combination on QoS, which in turn assists in choosing better arms, i.e., combination of entries in successive rounds. 

In the next subsection, we first demonstrate the operation of AHP considering static weights in PCM. Thereafter, the operation of CAMAB is described in choosing the PCM weights dynamically based on environmental feedbacks. 

\subsection{Operation of AHP}
The operation of AHP to compute the ranking between modes for all three slices is described below. 
\subsubsection{Ranking for eMBB applications} 
\paragraph{Comparison between criteria at Level 0}
\noindent The eMBB applications require highest data rate. Therefore, data rate (C1) is given the highest priority (i.e., $10$) while constructing the pairwise comparison matrix (PCM), among other criteria (shown in Table \ref{pmcembb}). The other three criteria, i.e., reliability (C2), latency (C3) and jitter (C4) have been assigned the weights $5$, $2$ and $1$ respectively.

\begin{table}[]
    \centering
     \caption{Initial criteria weights for different slices}
    \begin{tabular}{|c|c|c|c|}
    \hline
    \multirow{2}{*}{\textbf{Criteria}} & \multicolumn{3}{c|}{\textbf{Weights}}\\\cline{2-4}
    & eMBB & uRLLC &mMTc\\\hline
    Datarate (C1) & 10 &3 &1\\
    Reliability (C2) & 5 &6 &3\\
    Latency (C3) & 3 & 7 &6\\
    Jitter (C4) &4 &1 &9\\\hline
    \end{tabular}
    \label{cr_weights}
\end{table}

The matrix PCM is a $4 \times 4$ real matrix, where the entry $x_{ij}$ represents the relative importance of $i^{th}$ criterion as compared to the $j^{th}$ criterion. Therein, $x_{ij}>1$ implies that the $i$-th criterion is more important than the $j$-th one; $x_{ij}<1$ implies that the $i$-th criterion is less important than the $j$-th one; whereas $x_{ij}=1$ represents that the $i$-th and $j$-th criteria are equally important. Here, $i, j \in \{1, 2, 3, 4\}$.

Based on PCM, the normalized pairwise comparison matrix (NCM) is computed. The matrix \texttt{Y} is also a $4 \times 4$ real matrix, where  $y_{ij}$ is computed as: $y_{ij}=\frac{x_{ij}}{\displaystyle \sum_{j=1}^4 x_{ij}}$. The criteria weight (W) is computed by averaging of each row of the NCM (last column of NCM). Each element in W ($4 \times 1$ real matrix) is computed as: $w_{i}={\displaystyle \sum_{j=1}^4 \frac{y_{ij}}{4}}$.
 
In the next step, based on the PCM and the NCM, the consistency ratio (CR) is computed as follows. First, the principle eigen value  $\lambda_{max}$ is computed by multiplying each element of the column vector W (in NCM) by the corresponding elements of the row vector obtained by summing up the elements of the PCM in a column wise fashion. In this case, $\lambda_{max}$ is $3.8$.  Next, we compute the consistency index (CI) as: CI$=\frac{\lambda_{max}-n}{n-1}$, where $n(=4)$ is the total number of criteria. Finally, consistency ratio (CR) is computed to check the consistency of the pairwise evaluations made in the PCM. The value of $CR$ is computed as $CR$=$\frac{\text{CI}}{\text{RI}}$, where RI is ratio index. RI=$0.90$ when $n=4$. The sign of CR signifies the consistency \cite{SAATY1977234}. A positive CR implies that the NCM is  consistent. A negative CR with large absolute value signifies high inconsistency. However, when the absolute value of a negative CR is small, it means that the inconsistency is minor and the NCM configuration is acceptable for practical usage. In this case, CR (in the NCM) is $-0.07$, which is \emph{less than} $0.1$, showing that the pairwise evaluations are \emph{consistent}. 

\paragraph{Comparison between the modes at Level 1}
In this section, first we construct the PCMs between the modes for each of the criteria, i.e., for data rate, reliability, latency and for jitter as criterion. We follow the same procedure described in the previous section to construct these PCMs. Subsequently, based on the PCMs, we compute the NCMs for all the criteria. The NCMs for data rate, reliability, latency and jitter are depicted in Table \ref{pmcembb}. The consistency ratios suggest that the PCMs are consistent.

\paragraph{Rank computation}
To compute the ranks for different modes in the context of eMBB services, first we construct matrix \texttt{P} of order $4\times4$ containing the $4 \times 1$ vectors of criteria weights, i.e., $CW_1$, $CW_2$, $CW_3$ and $CW_4$. Then, we compute the rank by multiplying the matrix \texttt{P} by the column vector W which is a $4 \times 1$ matrix.  It can be inferred from the ranking that NR and D2D communications are better options as compared to LTE-A, as far as eMBB applications are concerned.

\begin{table*}[t]
\centering
\renewcommand{\arraystretch}{1.2} 
\caption{PCM and NCM for different criteria and modes\\}
\resizebox{\textwidth}{!}{%
\begin{tabular}{|c|c|ccccc|ccccc|}
\hline
\multirow{2}{*}{\textbf{Level of AHP}} & \multicolumn{1}{c|}{\textbf{Criteria}/} & \multicolumn{5}{c|}{\textbf{PCM}} & \multicolumn{5}{c|}{\textbf{NCM}} \\ \cline{3-12}
 & \textbf{Options} &  & C1 & C2 & C3 & C4 & C1 & C2 & C3 & C4 & W\\ \hline

\multirow{13}{*}{\textbf{Level $0$}} 
 & Pairwise comparison & C1 & 1  & 2 & 5 & 10 & 0.556 & 0.571  & 0.526 & 0.556 & 0.55 \\
 & between criteria & C2 & 0.5 & 1 & 3 & 5 &  0.278 & 0.286 & 0.316 & 0.278 & 0.28 \\
 & eMBB & C3 &0.2  & 0.3 & 1 & 2 &  0.111 & 0.086 & 0.105 & 0.111 & 0.1  \\
 & $\lambda_{max}=3.98$, CR$=-0.0006 < 0.1$ & C4 & 0.1  & 0.2 & 0.5 & 1 &0.056 & 0.057  & 0.053 & 0.556 & 0.05 \\\cline{2-12}

 & uRLLC, & C1 & 1  & 0.5 & 0.42 & 3 & 0.179 & 0.177  & 0.174 & 0.176 & 0.176 \\
  & $\lambda_{max}=3.94$,   &  C2 & 2 & 1 & 0.85 & 6 & 0.357 & 0.355 & 0.353 & 0.353 & 0.354 \\
   & CR$=-0.01 < 0.1$ &C3 & 2.3  & 1.16 & 1 & 7 & 0.411 & 0.411 & 0.415 & 0.412 & 0.412 \\
    & &  C4 & 0.3 & 0.16 & 0.14 & 1&0.054 & 0.057  & 0.058 & 0.059 & 0.056  \\\cline{2-12}

 & mMTC & C1 & 1  & 0.3 & 0.16 & 0.1 &0.053 & 0.048  & 0.059 & 0.05 & 0.052 \\
  & $\lambda_{max}=3.64$, &  C2 & 3 & 1 & 0.5 & 0.3 & 0.158 & 0.159 & 0.018 & 0.15 & 0.121  \\
   & $\text{CR}=-0.132 < 0.1$ & C3 & 6  & 2 & 1 & 0.6 & 0.316 & 0.317 & 0.369 & 0.300 & 0.325 \\
    &  & C4 & 9 & 3 & 1.5 & 1 & 0.474 & 0.476 & 0.554 & 0.5 & 0.5  \\\hline


\multirow{16}{*}{ \textbf{Level $1$} } & & & LTE & NR & D2D & & & LTE & NR & D2D & $CW_i$\\\hline
& Data rate & LTE & 1& 0.33 & 0.5 & & LTE & 0.3 & 0.3 & 0.3& 0.3  \\
& $\lambda_{max}=3.007$ & NR &3 &1& 2 & & NR& 0.3 & 0.3 & 0.3& 0.3   \\
& CR$=0.0065 < 0.1$  & D2D & 2  & 0.5 & 1 & &D2D& 0.39  & 0.39 & 0.4 & 0.39   \\\cline{2-12}
& Reliability& LTE & 1 & 0.5 & 0.33 & & LTE & 0.52 & 0.14 & 0.18 & 0.28  \\
&$\lambda_{max}=2.336$  & NR & 0.6 & 1 & 0.5 & & NR& 0.316 & 0.28 & 0.27 & 0.29   \\
& $\text{CR}=-0.57 < 0.1$) & D2D & 3  & 2 & 1  & & D2D& 0.15  & 0.57 & 0.54 & 0.42\\\cline{2-12}
&Latency& LTE & 1 & 0.5 & 0.25 & & LTE & 0.143 & 0.143 & 0.161 & 0.149  \\
& $\lambda_{max}=2.859$, & NR & 2 & 1 & 0.3 & & NR& 0.186 & 0.186 & 0.191 & 0.254   \\
&$\text{CR}=-0.1212 < 0.1$ & D2D & 4  & 2 & 1  & & D2D& 0.571  & 0.571 & 0.645 & 0.596   \\\cline{2-12}
& Jitter & LTE & 1 & 2 & 0.75 & & LTE & 0.333 & 0.364 & 0.36 & 0.35  \\
& $\lambda_{max}=2.9140$ & NR & 0.5 & 1 & 0.3 & & NR& 0.167 & 0.182 & 0.146 & 0.16  \\
& CR$=-0.074<0.1$ & D2D &1.5 & 2.5 & 1  & & D2D& 0.5  & 0.45 & 0.48  & 0.48   \\\hline

\end{tabular}%
}
\label{pmcembb}
\end{table*}

\subsubsection{Ranking for uRLLc services}
To compute the rank between the modes towards serving uRLLc services, we follow a method similar to that described for eMBB services. For uRLLc services, latency and reliability have been given higher weights than reliability and jitter. Accordingly, we compute PCM and NCM. Then, the ranking between the modes are computed by multiplying matrix P=[$CW_1$, $CW_2$, $CW_3$, $CW_4$] (computed in the previous section) with the $4 \times 1$ column vector W as computed for uRLLc. CR values suggest that the pairwise comparisons made is consistent. The ranking infers that D2D and NR communication modes are preferred as compared to LTE, as far as uRLLc is concerned.

\begin{table}[t]
\centering
\renewcommand{\arraystretch}{1.2} 
\caption{Final ranking\\}
\begin{tabular}{|c|ccc|ccc|}
\hline
\multirow{2}{*}{\textbf{Options}} & \multicolumn{3}{c|}{\textbf{Scores}}&\multicolumn{3}{c|}{\textbf{Rank}}\\\cline{2-7}
& eMBB & uRLLC &mMTc &eMBB & uRLLC &mMTc\\\hline

LTE & 0.199 & 0.205 &0.262 & 3 &3 &2\\
NR & 0.41 & 0.27 & 0.224 & 1&2&3\\
D2D &0.24 & 0.47 & 0.495 & 2 &1 &1\\\hline
\end{tabular}
\label{ranking}
\end{table}

\subsubsection{Ranking for mMTC services}
To rank the modes in serving mMTC services, jitter and latency have been given highest weights ($9$ and $6$ respectively). Accordingly, we compute PCM and NCM. Finally, we compute the rank by multiplying matrix P with the $4 \times 1$ column vector W as computed for mMTc. CR values suggest that the pairwise comparisons made in Table \ref{pmcembb} are \emph{consistent}. The rank shows that LTE and D2D communications are more suitable than NR, as far as mMTC services are considered. The final rank computations for eMBB, uRLLc and mMTc slices are shown in Table \ref{ranking}.

\subsubsection{The mode selection algorithm}
The rank computations at time $t$ (shown in Table \ref{ranking}) may not hold good at time time $t+\Delta$ due to channel fading, presence of dynamic obstacles and UE mobility. To account the aforementioned factors, we combine the dynamic variation of RSRP from different modes with the static AHP based ranking. To normalize the RSRP values received from different networks, we employ the Sigmoid function which maps any real valued number into a value between 0 and 1. Then, based on the AHP based rank, we compute the dynamic rank by multiplying the $3 \times 1$ matrices (consisting of scores) by the $1 \times 3$ vector, consisting of the values of $\frac{1}{1+e^{-x_i}}$ where $x_i$ is the RSRP from corresponding options (shown in Table \ref{ranking}). 

\subsection{Operation of CAMAB to compute the Level 0 PCM of AHP}\label{mab}

Depending on the changing channel conditions, user service requirements and UE mobility patterns, the PCM at Level 0 of the AHP needs to be adjusted dynamically. This can be achieved by sequentially learning the optimal configuration of the PCM that maximizes the quality of experience for the desired criteria. To dynamically configure the PCM, we utilize the CAMAB framework \cite{CAMAB}.  Combinatorial MABs are designed specifically to handle combination of choices and keep regret to a minimum. The adversarial nature of the proposed CAMAB is best suited to handle the nonstationarity in the network environment that arises due to random UE mobility and ever changing QoS requirement thresholds of the UEs with time. Adversarial bandits refer to a class of Multi-armed bandit problems where the reward sequence is generated by an adversary (in this case the dynamic network environment). At each time step, the agent follows a \emph{policy} that maps past observations and corresponding actions to a probability distribution over arm choices, and thereafter selects an arm accordingly. After selecting that arm, the agent receives a reward. The objective is to minimize regret \footnote{Difference between the optimal reward and the reward from the chosen arm.} with respect to the best arm. This adversarial bandit framework is extended by CAMAB where the agent selects a subset of arms (i.e., a combinatorial action), rather than a single arm at each time step. The reward is typically a function of the selected subset of arms. This setting requires specialized algorithms to efficiently explore a combinatorial large action space while maintaining low regret.  

\paragraph{CAMAB framework} Let $PCM_{ij}$ be the entry in the $i^{th}$ row and $j^{th}$ column of the Level 0 PCM. Our proposed CAMAB framework aims to choose the values of upper diagonal entries (i.e., $i < j$) of PCM. It may be observed that $PCM_{ii}=1$ and $PCM_{ji}=\frac{1}{PCM_{ij}},\quad \forall i, j$. There are 6 entries in the upper diagonal of PCM: $PCM_{12}$, $PCM_{13}$, $PCM_{14}$, $PCM_{23}$, $PCM_{24}$ and $PCM_{34}$. We need to choose a combination of these 6 entries using the proposed CAMAB. It may be observed from Table S1 that each $PCM_{ij}$ has a total of 17 total possible choices, i.e., $\{x, \frac{1}{x} | x=1,\; 2,\; 3,\; 4,\; 5,\; 6,\; 7,\; 8,\; 9\}$. With partial human feedback, the total number of such choices for each entry can be further reduced. 

To illustrate the CAMAB framework, let us consider that the number of possible choices for $PCM_{12}$, $PCM_{13}$, $PCM_{14}$, $PCM_{23}$, $PCM_{24}$ and $PCM_{34}$ are $K_1$, $K_2$, $K_3$, $K_4$, $K_5$ and $K_6$ respectively (after considering human feedback). Clearly, $K_i \leq 17, \; \forall i \in [6]$.

\begin{definition}{\textbf{Base Arm}}
A possible choice of priority for a particular $PCM_{ij}$ is called a base arm.
\end{definition}

\noindent Hence, the number of base arms for $PCM_{12}$, $PCM_{13}$, $PCM_{14}$, $PCM_{23}$, $PCM_{24}$ and $PCM_{34}$ are $K_1$, $K_2$, $K_3$, $K_4$, $K_5$ and $K_6$ respectively. Therefore, total number of base arms is $\displaystyle\sum_{i=1}^6K_i$. The objective of the CAMAB is to choose a combination of $6$ base arms out of $\displaystyle\sum_{i=1}^6K_i$ possible base arms. In this case, a combination can be represented using a {\it binary vector} of dimension $d = \displaystyle\sum_{i=1}^6K_i$, which is organized into six consecutive blocks as follows:

\begin{itemize}
    \item The first $K_1$ components indexed by $V(1), V(2), \ldots, V(K_1)$ correspond to the $K_1$ possible alternatives for $PCM_{12}$. Here $V(i)=1$ if base arm $i$ is chosen for $PCM_{12}$; Otherwise $V(i)=0 \; \forall i\in [K_1]$.
    
    \item The next $K_2$ components indexed by $V(K_1+1), V(K_1+2), \ldots, V(K_1+K_2)$ correspond to the $K_2$ possible alternatives for $PCM_{13}$. Here also, $V(i)=1$ if base arm $i$ is chosen for $PCM_{13}$; Otherwise $V(i)=0; \; \forall i \in \left\lbrace K_1+1,\; K_1+2,\; \ldots,\; K_1+K_2  \right\rbrace$.
    
    \item This pattern continues similarly for the consecutive $K_3$, $K_4$, $K_5$ and $K_6$ components of the vector corresponding to possible alternatives for $PCM_{14}$, $PCM_{23}$, $PCM_{24}$ and $PCM_{34}$ respectively.
\end{itemize}

At each time instant $t$ the CAMAB agent chooses a binary vector and receives a reward for its choice. Each such binary vector is referred as \emph{arm}.  Subsequently, the arm chosen at time $t$ is denoted as $A(t)$; and the set of all possible arms of the CAMAB is denoted by $\mathcal{A}$.

\paragraph{Reward design} At time step $t$, let the CAMAB agent choose arm $A(t)\in\mathcal{A}$. Upon choosing, the agent receives a reward corresponding to its choice. Over time, this enables the agent to identify the optimal arm such that the resulting priority assignment aligns with the service requirements of the UE. {\it The reward of an arm is the summation of rewards obtained for choosing the corresponding base arms.}

To illustrate, let us consider that the QoS threshold for the $j^{th}$ criteria is $\bar\mu_j$ where, $j$ is 1 for datarate, 2 for reliability, 3 for latency and 4 for jitter. Moreover, to capture overhead due to handover, we consider the $\bar\mu_h$ as the QoS threshold for handover rate. After choosing $A(t)$, a mode is selected for the UE using the AHP process. Let $\mu_j$ be the actual measurements of QoS criteria $j$ after mode selection and $\mu_h(t)$ be the current handover rate at time step $t$. The total reward for choosing $A(t)$  is computed as:
\begin{equation}
  R(t)=\begin{cases}
      \displaystyle\sum_{i,j=1\; i<j}^4 R_{ij}(t)+\frac{1}{1+e^{\mu_h-\bar\mu_h}}, & {\rm consistent\;PCM}\\
      -L, & {\rm otherwise}
  \end{cases}  
\end{equation}
\noindent where $R_{ij}$s correspond to the $PCM_{ij}$ entries, and are computed as follows:
\begin{equation}
    R_{ij}(t)=\begin{cases}
        S(\mu_{i}-\bar\mu_i)+ S(\mu_j-\bar\mu_j), & i=1,\;j=2.\\
        S(\mu_i-\bar\mu_i)+\bar S(\mu_j-\bar\mu_j), & i=1,2; \;j=3,4.\\ 
        \bar S(\mu_i-\bar\mu_i) +\bar S(\mu_j-\bar\mu_j), & i=3,\;j=4.
    \end{cases}
\end{equation}

\noindent Here, $S(x)=\frac{1}{1+e^{-x}}$ and $\bar S(x)=\frac{1}{1+e^{x}}$. Here, the function $S$ is used for the criteria to be maximized (data rate and reliability) and the function $\bar S(x)$ is used for the criteria to be minimized (latency and jitter). \\

\paragraph{Computing optimal policy}
To compute the optimal policy, i.e., the optimal probability distribution of choosing arms, we adopt algorithmic framework of CAMAB as proposed in \cite{CAMAB}. We begin by embedding the action set $\mathcal{A}$ into the $d$ dimensional simplex, namely $\mathcal{P}$. Therein, $A(t)$s are divided by the maximum number of base arms chosen in a combination (i.e., $6$). The algorithm accepts $\mu^0$, $T$, $d$, $\lambda$ and $\mu_{min}$ as input parameters, where  $\mu^0$ represents an uniform probability distribution across the arms, $T$ represents the count of total number of time steps elapsed since the beginning of the algorithm, $d$ represents the dimension of the action vector, $\lambda$ represents the smallest nonzero eigenvalue of $\mathbb{E}[A(t)A^\top(t)]$ and $\mu_{min}$ is the smallest component of the exploration-inducing probability distribution vector. 

\begin{definition}{$\mu_{min}$}
 $A_i(t)$ is the $i^{th}$ component of the action $A(t)$, belonging to the set $\mathcal{A}$. The total number of arms in set $\mathcal{A}$ is $\displaystyle\prod_{j=1}^6 K_j$. Accordingly, $\mu_{min}$ is defined as $\displaystyle \min_{i} (6 \times \mu^0_{i})$, where, $\mu^0_{i}$ is the $i$-th component of the vector $\mu^0$, representing an uniform probability distribution across the arms. Here $\mu^0_{i}=\frac{1}{6\displaystyle\prod_{j=1}^6K_j}\displaystyle\sum_{A(t)\in \mathcal{A}}A_{i}(t),\quad \forall i\in[d]$. 
\end{definition}
We initialize $T$=$1$, $q^0$=$\mu^0$, $\eta=\gamma C$ and $C=\frac{\underline{\lambda}}{6^{3/2}}$. At each timestep \(t\), we compute $\gamma=\frac{\sqrt{6\log \mu_{\min}^{-1}}}{\sqrt{6\log \mu_{\min}^{-1}}+\sqrt{C(Cm^2d+6)T}}$ and a smoothed distribution \(q'(t-1)\) by mixing \(q(t-1)\) with \(\mu^0\) (Step 3). Thereafter, a distribution \(p_{A(t)}(t-1)\) is computed over \(\mathcal{A}\) whose mean vector equals \(6q'(t-1)\) (step 4). 
Next, an action \(A(t)\) is sampled according to \(p_{A(t)}(t-1)\) and the reward \(R(t)\) is observed (step 5). The observed reward is projected onto the action space using the pseudo-inverse of the second-moment matrix \(\Sigma(t-1)=\mathbb{E}[A(t-1)A^\top(t-1)]\) of the chosen distribution, yielding an unbiased estimate \(\tilde{R}(t)\) (steps 6 and 7). Finally, the distribution \(q(t)\) is updated using exponential gradient with learning rate \(\eta = \gamma C\), and followed by a Bregman\footnote{A Bregman projection is a generalization of the Euclidean projection, used to map a point onto a closed convex set by minimizing a Bregman divergence rather than the standard Euclidean distance.} projection onto \(\mathcal{P}\) with respect to the KL divergence\footnote{Kullback-Leibler (KL) divergence is a non-symmetric measure of the difference between two probability distributions.} (Steps 8 and 9).  The total regret of the proposed algorithm upto time $T$ is $O(\sqrt{T})$ \cite{CAMAB}. This is sublinear in time $T$ and hence the algorithm converges to the optimum. The pseudocode of the algorithm is provided in Algorithm \ref{camab}. 

\begin{algorithm}[h]
    \small
    \caption{Policy computation using CAMAB}
    \label{camab}
    
    \SetKwInput{KwInit}{Input}
    
    \KwInit{
        $\mu^0$, $T$, $d$, $\lambda$ and $\mu_{min}$.
    }
    Set $q_0 = \mu^0$, $T = 1$,  $C = \frac{\underline{\lambda}}{6^{3/2}}$ and $\eta = \gamma C$\;
    \For{$t \ge 1$}{
    Compute $\gamma = \frac{\sqrt{6\log \mu_{\min}^{-1}}}{\sqrt{6\log \mu_{\min}^{-1}} + \sqrt{C(36Cd+6)T}}$ and
        $q'(t-1) = (1-\gamma)q(t-1) + \gamma\mu^0$\;
        Select a distribution $p(t-1)$ over $\mathcal{A}$ such that $\displaystyle\sum_{A(t)\in\mathcal{A}} p_{A(t)}(t-1) A(t) = 6q'(t-1)$\;
        
        Select a random arm $A(t)$ with distribution $p_{A(t)}(t-1)$ and get reward $R(t)$\;
        
        $\Sigma(t-1) = \mathbb{E}[A(t-1)A^{\top}(t-1)]$\;
        
        $\tilde{R}(t) = R(t)\Sigma(t-1)^{+}A(t)$, where $\Sigma(t-1)^{+}$ is the pseudo-inverse of $\Sigma(t-1)$\;
        
         $\tilde{q}_i(t)= q_{i}(t-1) \exp(\eta \tilde{R}_{i}(t)), \quad \forall i \in [d]$\;
        
        Compute $q(t)$ as the projection of $\tilde{q}(t)$ onto the set $\mathcal{P}$ using the KL divergence\;
        
        $T = T + 1$\;
    }
\end{algorithm}

\section{Results and discussions} \label{results}
In this section, we present the performance studies on our proposed mode selection approach and compared with three existing mode selection approaches: an RSRP-based \cite{RSRP_baseline}, a Software Defined Networking based (SDN\_joint) \cite{b7}, and a joint mode selection, relay selection and resource allocation (JMSRA) \cite{JMSRA} approaches. All UEs in a run share the same slice (eMBB/uRLLc/mMTc). In \cite{RSRP_baseline}, a network selection algorithm has been proposed for D2D network utilizing RSRP and location information of the base stations. In \cite{b7}, a handover mechanism has been proposed for a D2D enabled SDN. Therein, half handover is executed if the handover condition is satisfied by only one UE, but full handover is executed if the handover condition is satisfied by both the UEs. In \cite{JMSRA}, mode selection, relay selection and resource allocation problems for D2D have been formulated jointly using a binary integer non-linear programming problem (BINLP). Therein, the QoS has been modeled by a threshold on data rate without any consideration to delay, jitter and packet loss. In the next subsection, the simulation set-up is described.

\subsection{Simulation set-up}
We consider a 1000 $\times$ 1000 square meters area with randomly placed UEs and randomly deployed Evolved Node Bs (eNBs) and Next Generation Node B (gNBs). Locations of eNBs and gNBs are chosen uniformly within the area at the start of each simulation run, while D2D links are formed by randomly pairing UEs. Mobility follows a simple step-based random waypoint model. We sweep UE speed from the set $\{2,4,6,8,10\}\,\si{m/s}$ and, the number of concurrent UEs from the set $\{20,40,60,80,100\}$. The UE speeds are considered as fixed.

The physical layer is modeled via an Low-Density Parity-Check (LDPC) coded link over Additive white Gaussian Noise (AWGN) with
path-loss and log-normal shadowing. For each UE and each candidate
technology (NR, LTE-A and D2D), the RSRP is computed based on 3GPP Urban Macro (UMa) propagation model \cite{uma3gpp}. Thereafter, RSRP is converted
to Signal-to-noise ratio (SNR) using the thermal noise and technology-specific noise figures. The PHY simulator then applies modulation-dependent SNR penalties, Doppler shift and speed penalties, and LDPC coding to obtain block error rate (BLER) and effective throughput. In the simulation environment, Hybrid automatic repeat request (HARQ) and scheduling delays are modeled using deterministic Transmission Time Interval (TTI) durations and speed/load-dependent queuing delays; the overall latency and jitter (short-term latency variance) are computed on per packet basis.

\subsubsection*{Key Parameterization}
Table~\ref{tab:sim-params} summarize the main simulation, selection, and PHY parameters. Slice weights bias the QoS utility function towards different Radio Access Technologies (eMBB: throughput-centric; URLLC: reliability/latency; mMTC: high connection density). Handover (HO) hysteresis thresholds are set per target technology to mitigate ping-pong.

\begin{table}[t]
  \centering
  \caption{Global simulation parameters.}
  \label{tab:sim-params}
  \begin{tabular}{ll}
    \hline
    Parameter & Value \\ \hline
    Simulation area & $1000 \times 1000~\mathrm{m}^2$ \\
    No.\ of 5G gNBs / LTE eNBs & $2 / 1$ \\
    Carrier frequency (5G / LTE / D2D) 
      & $5.5 / 2.1 / 2.4~\mathrm{GHz}$ \\
    System bandwidth (per link) & $20~\mathrm{MHz}$ \\
    Subcarrier spacing (5G / LTE / D2D) 
      & $30 / 15 / 15~\mathrm{kHz}$ \\
    BS transmit power (5G / LTE) & $35~\mathrm{dBm}$ \\
    D2D transmit power & $15~\mathrm{dBm}$ \\
    HO hysteresis & $0.6$~$\mathrm{dB}$ \\
    Noise figure (5G / LTE / D2D) 
     & $6 / 7 / 5~\mathrm{dB}$ \\
    Max D2D distance & 80 m \\
    User speeds & $\{2,4,6,8,10\}~\mathrm{m/s}$ \\
    No.\ of users & $\{20,40,60,80,100\}$ \\
    Packet size & $1500$~bytes \\
    Channel coding & LDPC, $n{=}512$, rate $\approx 1/2$ \\
    \hline
  \end{tabular}
\end{table}

All experiments were implemented in \textbf{Python} using \texttt{NumPy}, \texttt{SciPy}, \texttt{pyldpc}, \texttt{CommPy} and \texttt{Matplotlib} libraries. The simulation code has been made publicly available\footnote{Github link: \url{https://github.com/Avirup723/CAMAB-AHP}}. For each run, \textbf{base stations are placed uniformly at random} within the area. D2D links are ad-hoc (no fixed sites) and they are using mmWave frequency band. The presence of dynamic obstacles has been modeled as per 3GPP/ITU standard \cite{mmWaveRLF}. Within the simulation areas UEs follow a Random Waypoint mobility model. The PHY layer uses \textbf{LDPC} (pyldpc) over AWGN, with SNR including thermal noise, receiver NF, Doppler/ICI (speed-dependent), per-RAT diversity bonuses, and modulation-dependent penalties. Link adaptation selects an effective MCS bounded by RAT caps (5G up to 256-QAM; LTE/D2D up to 64-QAM). Bit error rate (BER) is measured after LDPC decoding; latency aggregates TX time (at achieved throughput), scheduling time, decoding time (SNR/MCS dependent), HARQ retransmission time (BLER-weighted), queuing (load \& speed sensitive), and propagation delays.

For each discrete point, the simulator advances several time steps; at each step we average per-UE metrics; then we average across steps to obtain one value per run; finally, we aggregate across repeated runs. We report the mean \(\bar{x}\) and a \(95\%\) confidence interval (CI) as follows:
\[
\bar{x} \;\pm\; 1.96\,\frac{s}{\sqrt{n}},
\]
where \(s\) is the sample standard deviation across runs and \(n\) is the number of repeated runs. The reported “error bar” is \(E = 1.96\,s/\sqrt{n}\). Confidence intervals report sampling uncertainty across repeated runs; where relevant, we also report the CI half-width \(E\) as the “error.”


\subsection{Convergence analysis of the proposed approach}
\begin{figure}
     \centering
     \includegraphics[width=8cm,height=4.5cm]{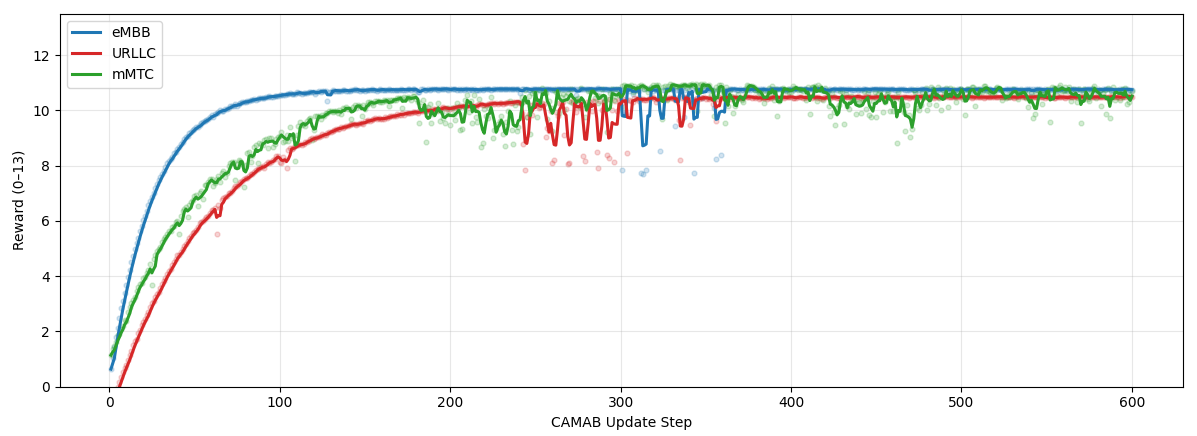}
    \caption{Slice wise convergence over episodes: eMBB slice converged around $200$ episodes, uRLLc converged around $400$ episodes and mMTc converged around 600 episodes.}
     \label{fig:conv_agents}
 \end{figure}

Figure \ref{fig:conv_agents} shows the convergence behaviors for different slices for the proposed approach. A window-averaged trend with window size of $5$ has been shown. It may be observed that all the slices exhibit steady state behavior at around $600$ episodes, where the eMBB slice converged the earliest followed by mMTc and uRLLc slices. The eMBB slice converged the fastest (around $200$ episodes), because the RL+AHP approach connects to optimal mode providing a strong and low-noise signal from the PHY layer at each step, ensuring high data rate (the key KPI). The \textbf{uRLLc} slice takes more steps to converge (around $400$ episodes), because the URLLC slice places the highest weight on reliability, a metric that is highly sensitive to channel fluctuations. The agent must therefore explore more widely and over more steps before identifying the optimal arm combination that consistently satisfies the tight reliability threshold. Finally, the mMTC slice takes the longest time to converge (around $600$ episodes) because the mMTC slice balances two parameters namely jitter and reliability which are highly susceptible to environmental changes and thus leading to higher degree of randomness, before the agent identifies optimal AHP weights.
\subsection{eMBB Slice: Maximizing throughput}

\begin{figure}
     \centering
     \includegraphics[width=9cm,height=6.5cm]{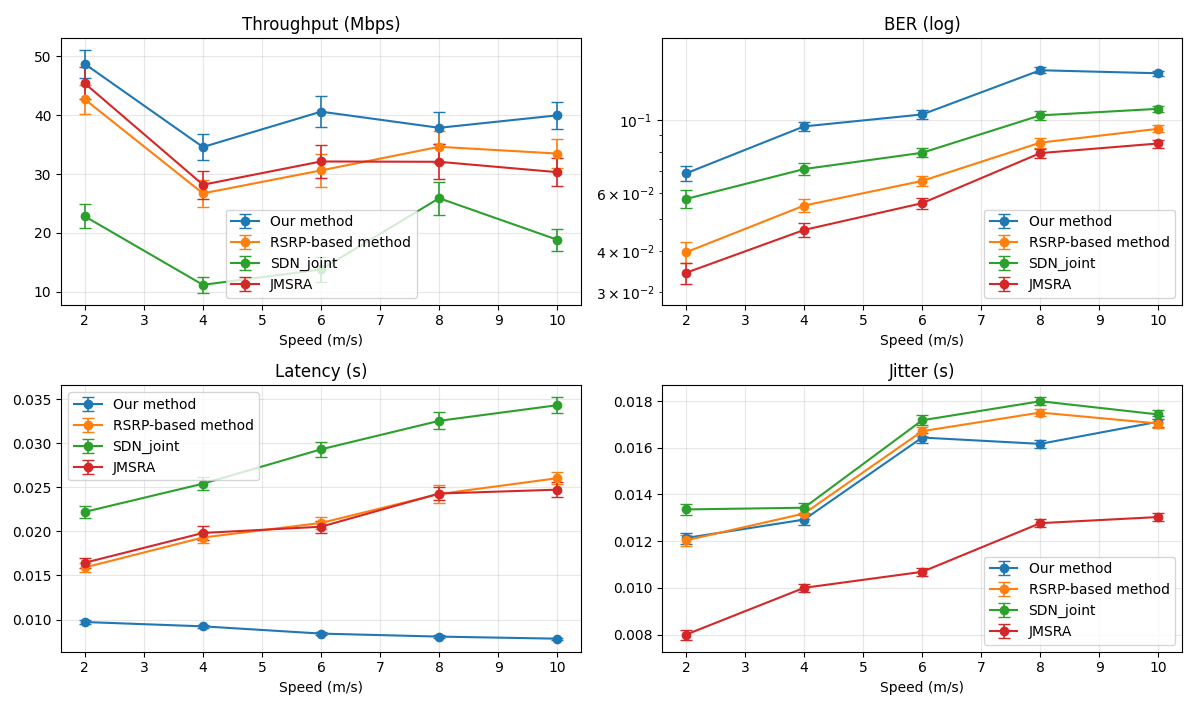}
    \caption{Performance evaluation for eMBB slice with varying speed showing that the proposed RL+AHP method outperforms the existing approaches in terms of throughput and latency.}
     \label{fig:embb_spd}
 \end{figure}

\begin{figure}
     \centering
     \includegraphics[width=9cm,height=6.5cm]{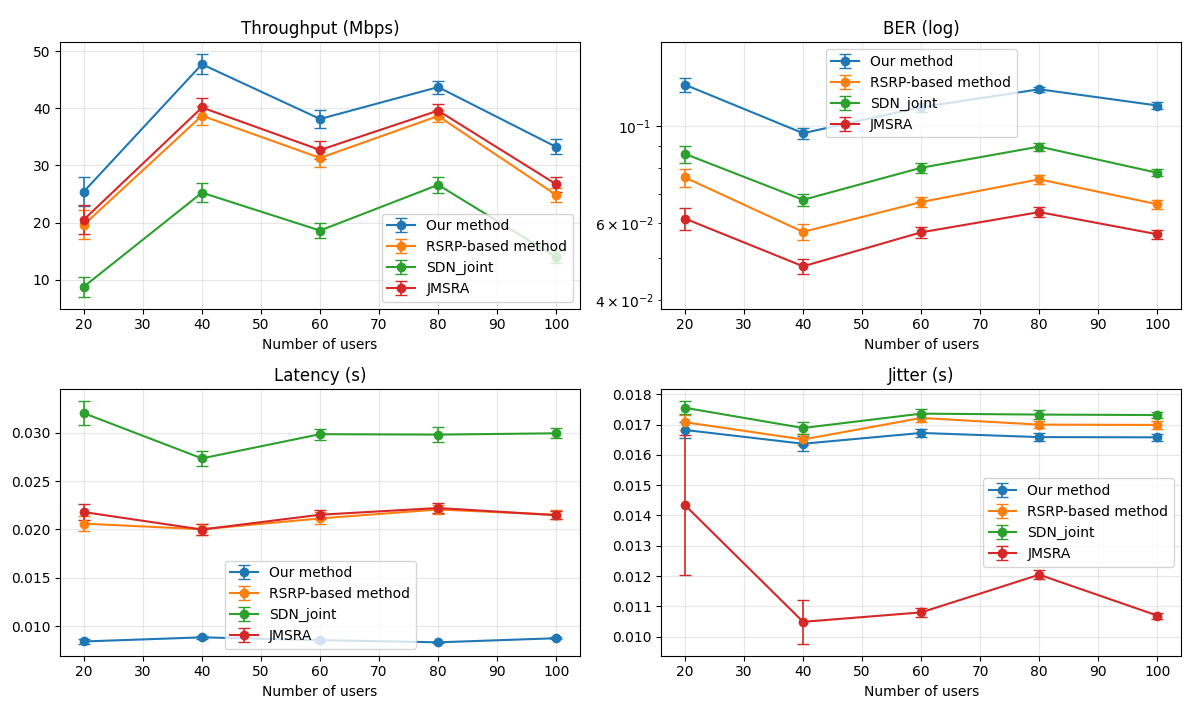}
    \caption{Performance evaluation for eMBB slice with varying number of users showing that the proposed RL+AHP method outperforms the existing approaches in terms of throughput and latency.}
     \label{fig:embb_user}
 \end{figure}

Figures~\ref{fig:embb_spd} and~\ref{fig:embb_user} report the results for eMBB slice. In both speed and user sweeps, \textbf{our method achieves the highest throughput (the primary KPI for eMBB) across all operating points}. In terms of throughput, our proposed RL+AHP method significantly outperforms the best performing existing work (i.e., JMSRA) with a maximum performance gain of $33\%$. The RSRP-based method achieves slightly less throughput, and SDN\_joint showing significantly degraded performance than JMSRA. This behavior is consistent because our method accounts both the QoS requirement along with non-stationarity of the environment while choosing the mode; thus it prefers 5G and D2D modes over LTE-A for low velocity users, whereas LTE-A and D2D for high velocity users leading to large throughput gain.

{\bf Our method attains the lowest latency for all the speed values ($2-10$ m/s)}, while all other approaches exhibit significantly high latency as compared to our approach. The SDN\_joint shows the largest latency because its fixed handover margin (HOM) and Time-to-trigger (TTT) thresholds discourage potentially beneficial HOs. The RSRP based and JMSRA approaches show similar performance in terms of latency because these approaches can not capture the impact of non-stationary caused by UE velocity, varying channel conditions and UE density. On the other hand, the proposed RL+AHP method outperforms all the related approaches because of its advanced QoS awareness and adaptability with the wireless environment. 

\begin{figure*}[t]
     \centering
     \begin{minipage}{0.32\textwidth}
     \includegraphics[width=5.5cm,height=4cm]{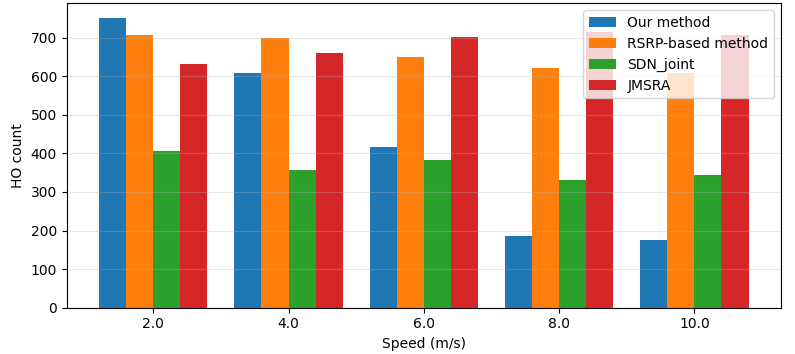}
    \caption{Evaluating total handover count for eMBB slice.}
     \label{fig:ho_embb}    
     \end{minipage}\hfill
     \begin{minipage}{0.32\textwidth}
\includegraphics[width=5.5cm,height=4cm]{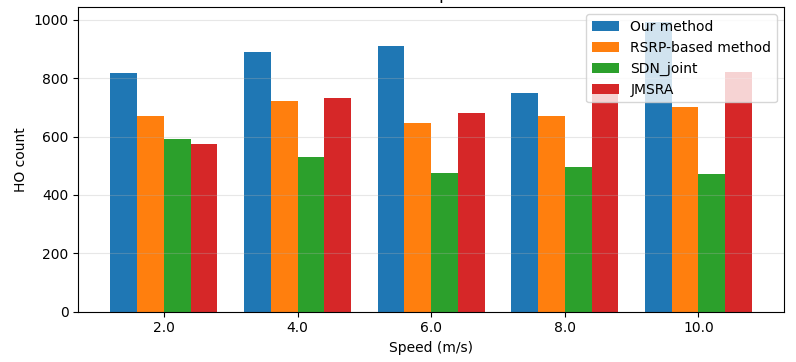}
    \caption{Evaluating total handover count for uRLLc slice}
     \label{fig:ho_urllc}
     \end{minipage}\hfill
     \begin{minipage}{0.32\textwidth}
\includegraphics[width=5.5cm,height=4cm]{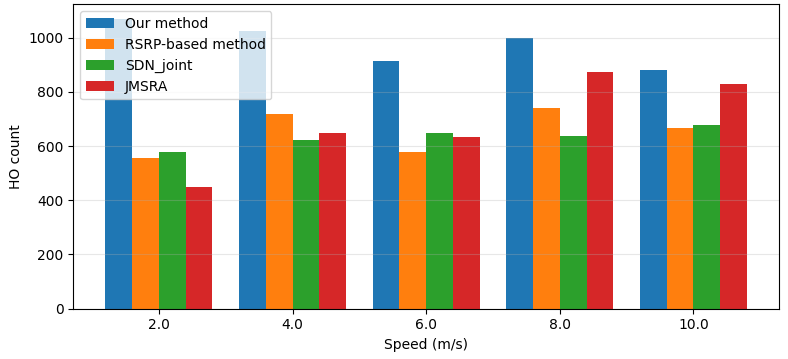}
    \caption{Evaluating total handover count for MMTC slice}
     \label{fig:ho_mmtc}  
     \end{minipage}  
 \end{figure*}

Fig.~\ref{fig:ho_embb} shows the handover count for varying UE speeds for all the approaches. The result shows that for extremely low velocity ($2$ m/s) the RL+AHP approach performs more handovers as compared to others. However, as the velocity increases ($4-10$ m/s) the RL+AHP approach starts outperforming the existing approaches (achieves the lowest handover count at $10$ m/s). This is because the RL+AHP approach prefers to connect with 5G and D2D modes for lower velocity in order to achieve higher throughput. However, as the velocity increases, the RL+AHP approach prefers to connect with the LTE-A mode to avoid unnecessary handovers, yet achieving the highest throughput and lowest latency. On the other hand, the other approaches aggressively handover to maintain lower BER and jitter, as evident from Figs ~\ref{fig:embb_spd} and~\ref{fig:embb_user}. Meanwhile, our proposed approach focuses on optimizing throughput and latency while maintaining the BER and jitter at acceptable levels, i.e., maintaining the required minimum thresholds. {\bf Thus, the RL+AHP approach strikes the best possible trade-off among other approaches for the eMBB slice.}

\subsection{uRLLc Slice: Ultra-Low Latency under strict reliability constraints}

\begin{figure}
     \centering
     \includegraphics[width=9cm,height=6.5cm]{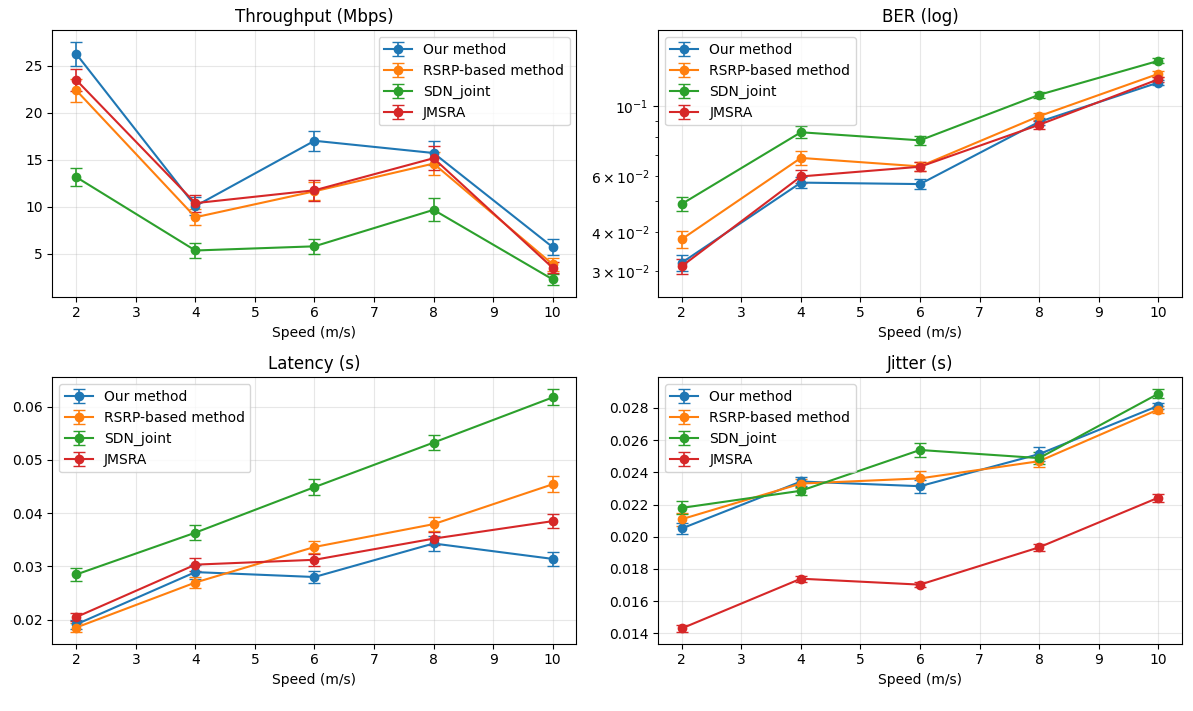}
    \caption{Performance evaluation for uRLLC slice with varying speed showing that the proposed RL+AHP method outperforms the existing approaches in terms of throughput, latency and BER.}
     \label{fig:urllc_spd}
 \end{figure}

\begin{figure}
     \centering
     \includegraphics[width=9cm,height=6.5cm]{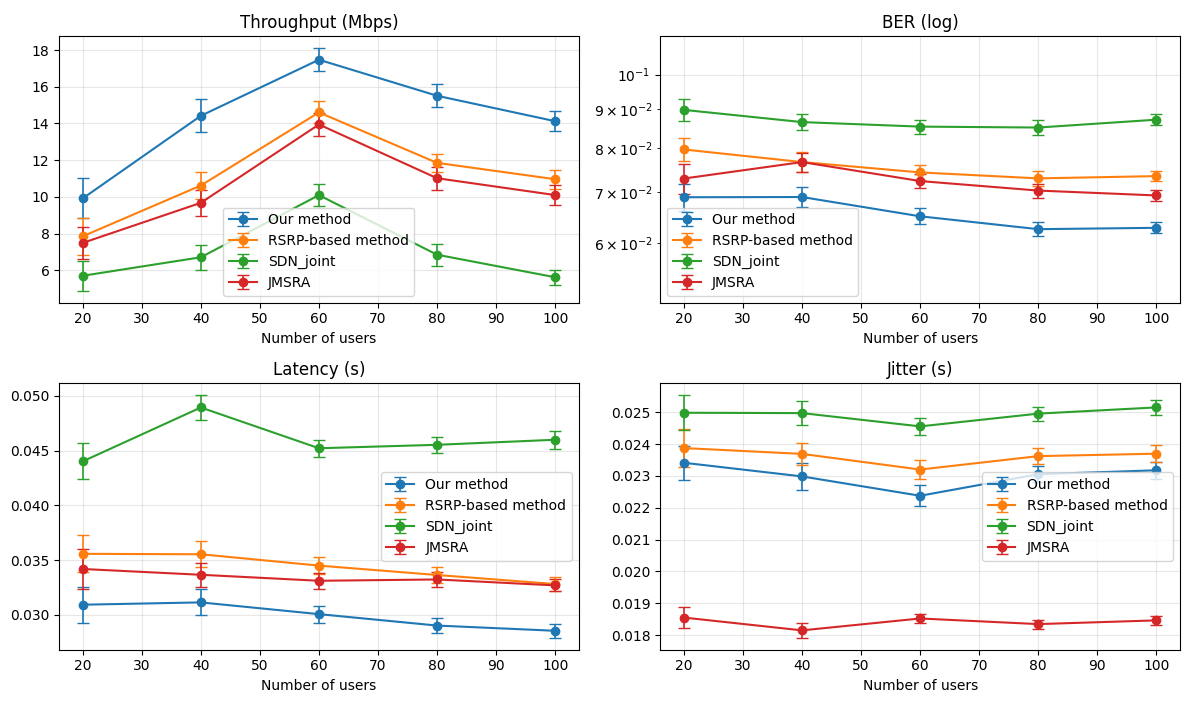}
    \caption{Performance evaluation for uRLLC slice with varying number of users showing that the proposed RL+AHP method outperforms the existing approaches in terms of throughput, latency and BER.}
     \label{fig:urllc_user}
 \end{figure}

Figures ~\ref{fig:urllc_spd} and ~\ref{fig:urllc_user} show the results for the uRLLc slice. {\bf The proposed RL+AHP achieves the lowest latency and BER (the primary KPIs for uRLLc) across all operating points.} In terms of latency and BER, our proposed RL+AHP outperforms the best existing work (i.e. JMSRA) with a maximum performance gain of $25\%$ and $10\%$ respectively. The performance of the RSRP-based method closely follows that of JMSRA, whereas the SDN\_joint exhibits the highest latency and BER. This behavior is consistent because our proposed RL+AHP approach is QoS aware and dynamically adapts to the non-stationary wireless environment. As a result, our method frequently prefers low-delay, highly reliable and high-SINR modes.

\textbf{Our proposed method outperforms the existing work in terms of throughput across all speed and user sweeps.} The RL+AHP method achieves a maximum throughput gain of $27\%$ against the best existing work (i.e. the RSRP-based approach). The JMSRA and RSRP based approaches show similar performance in terms of throughput, whereas the SDN\_joint approach has the lowest throughput because of its fixed handover thresholds which cannot adapt to a changing wireless environment to encourage beneficial HOs. It is also observed from Figure ~\ref{fig:urllc_user} that the throughput monotonically decreases after $60$ users. This is because interference increases with increasing number of users, thereby degrading the throughput.

Fig.~\ref{fig:ho_urllc} shows the handover count for varying UE speeds for all the approaches in the uRLLc slice. The RL+AHP approach performs more handovers as compared to the existing approaches across all values of user speed. This is because the RL+AHP approach focuses on minimizing the BER to adhere to the strict reliability constraints for the uRLLC slice. Therefore, the RL+AHP approach causes frequent handovers to cells that provide high SINR, thereby ensuring minimum BER and high reliability. The other approaches (e.g., SDN joint) reduce the handover count at the cost of degraded throughput, high latency and lower reliability which is considered unacceptable for uRLLc.

\textbf{From a uRLLc perspective, our method hits the best trade-off: lowest latency, lowest BER, high
throughput and moderate HO rate while maintaining the required threshold for jitter}, whereas JMSRA pushes reliability at the expense of mobility cost and SDN\_joint is too conservative in the context of
throughput and latency.

\subsection{mMTC Slice: high connection density}
\begin{figure}
     \centering
     \includegraphics[width=9cm,height=6.5cm]{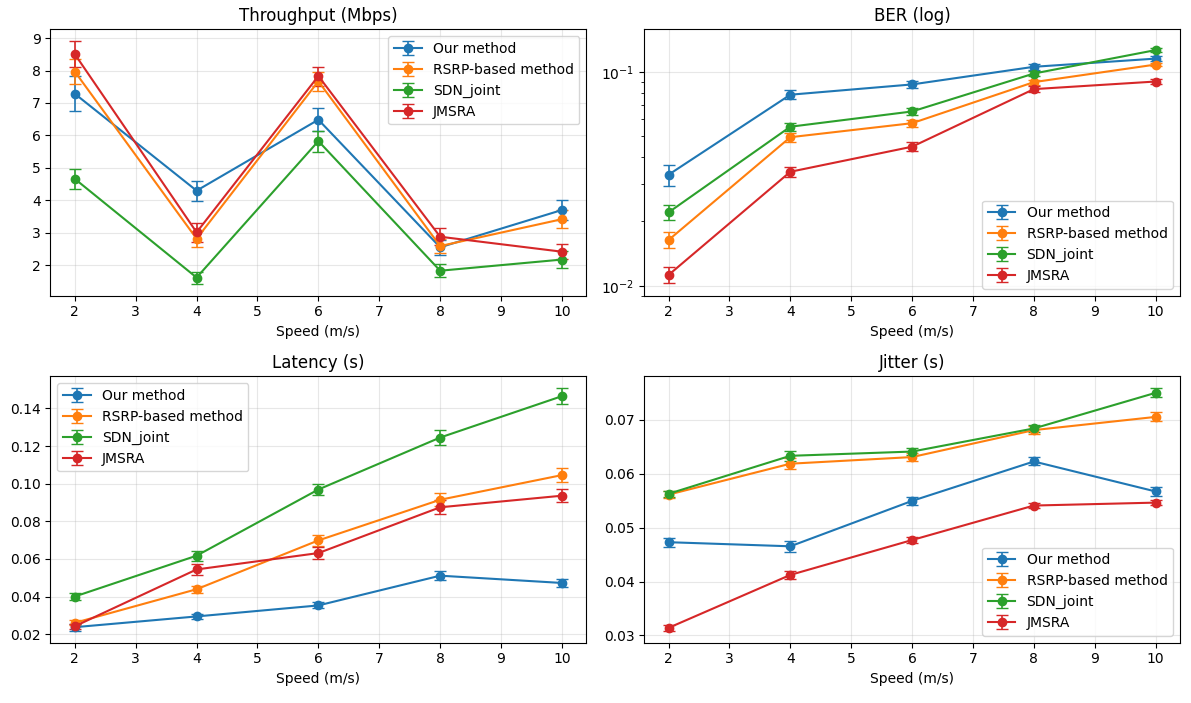}
    \caption{Comparison of mMTC users with varying speed}
     \label{fig:mmtc_spd}
 \end{figure}

\begin{figure}
     \centering
     \includegraphics[width=9cm,height=6.5cm]{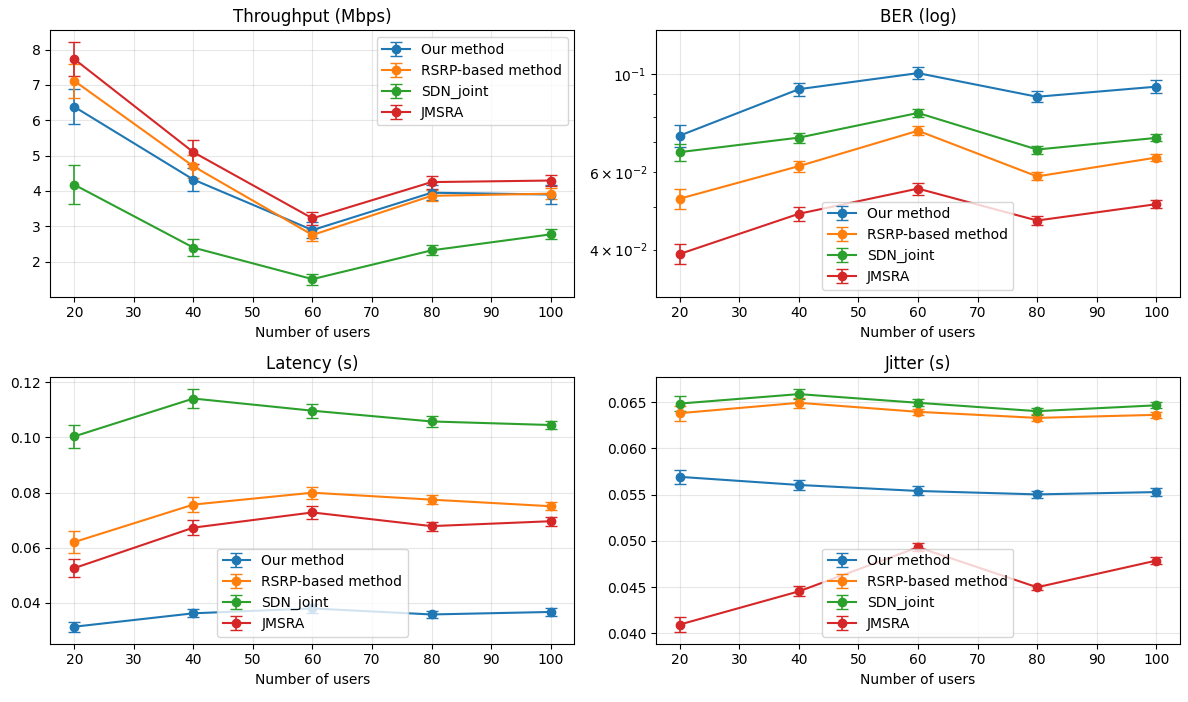}
    \caption{Comparison of mMTC users with varying number of users}
     \label{fig:mmtc_user}
 \end{figure}

The results for mMTC slice are reported in Figures ~\ref{fig:mmtc_spd} and ~\ref{fig:mmtc_user}. {\bf The proposed RL+AHP method outperforms existing approaches in terms of latency (a primary KPI for mMTC) across all operating points}. The RL+AHP method achieves a $45\%$ reduction in latency as compared to the best performing existing work (i.e. JMSRA). The RSRP-based approach follows the JMSRA closely, whereas the SDN\_joint approach suffers with the highest latency. On the other hand, the throughput for RL+AHP approach is lower than that for JMSRA; and the BER is relatively higher than the existing approaches. This can be explained as follows. The mMTC slice focuses on communication between massive number of densely deployed devices  for whom throughput requirements are modest, whereas latency and jitter are primary KPIs. Therefore, an mMTC user prioritizes modes that offer low latency along with minimal fluctuations in channel quality. As a result, an mMTC device avoid NR cells and mmWave D2D communications since the mmWave channel suffers from high propagation loss and fluctuations due to LoS blockage by dynamic obstacles. The mMTC user prefers connecting with a stable LTE-A cell or LTE-A assisted D2D over mmWave NR or mmWave D2D modes. Since the RL+AHP method is aware of the QoS requirements and can adapt with the wireless environment as opposed to the existing methods, it avoids the high bandwidth NR or NR assisted D2D modes, thereby sacrificing throughput and BER to significantly gain in latency as compared to the existing approaches. Our proposed approach also outperforms the SDN\_joint and RSRP-based approaches in terms of jitter, thereby establishing the fact that our approach prefers modes with stable channel conditions. It may be observed that the JMSRA marginally outperforms the RL+AHP approach in terms of jitter. However, our approach makes up for this by significantly outperforming JMSRA in terms of latency. \textbf{This indicates that the RL+AHP method utilizes slice aware ranking to accommodate additional devices without compromising connection stability}.

These results confirm that our proposed RL+AHP based slice-aware utility design is crucial: by
re-weighting throughput, latency, BER and HO cost per slice, our method
achieves a better QoS trade-off than the existing RSRP based,
SDN based and JMSRA strategies.

\subsection{Scalability of CAMAB}
\begin{figure}[h]
    \centering
    \includegraphics[scale=0.4]{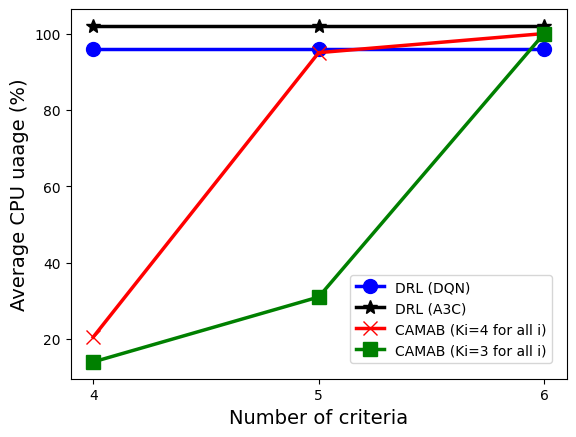}
    \caption{CPU usage vs. number of criteria.}
    \label{cpu_use}
\end{figure}

In this section, we compare the average CPU usage of the proposed RL+AHP method with two DRL models namely Deep Q Network (DQN) and Asynchronous Advantage Actor Critic (A3C) which are widely used to deal with the non-stationarity of the wireless environment \cite{1, 2, 3}. Fig. \ref{cpu_use} compares the average CPU usage of CAMAB based PCM configuration against that of DRL based mode selection for varying number of criteria. The CPU used for the experiment is an Intel Xeon processor operating at a base frequency of 2.20GHz. The setup consists of a single processor (one socket) that contains just one core. It has a 64-bit (x86\_64) architecture. The DQN is implemented as a standard feedforward neural network with 4 nodes each for the input and output layers and 64 nodes each for two hidden layers. The input layer corresponds to the dimension of the state vector, and the  4 nodes in the output layer correspond to the four actions as defined in Section \ref{probform}. The actor network in the A3C has the same architecture as the DQN. However, the critic network differs only in the output layer with just one node. It can be observed that, by selecting the $K_i$ values intelligently based on the QoS requirements of each application, the CPU usage of CAMAB is significantly reduced. Specifically, for $K_i = 4 ;\forall i \in {1, 2, 3, 4}$, CAMAB uses $4.7$ times less CPU than DQN and $5$ times less CPU than A3C. Furthermore, as the number of criteria increases, the CPU usage can be kept under control by further refining the values of $K_i$. For example, when the number of categories increases to $5$ the CPU usage can be reduced by setting $K_i=3$. This is an advantage that CAMAB based AHP has over DRL based approach. As the number of categories increases beyond $6$, the performance of DRL approaches become comparable with the proposed RL+AHP method. 

\section{Conclusion} \label{con}
We have proposed a mode selection algorithm for D2D enabled heterogeneous network by integrating a two-level AHP with a RL based method. The AHP part accounts the service requirements of application classes along with RSRP values of different modes to select the best possible option (i.e., LTE only, NR only, LTE via D2D and NR via D2D). On the other hand, the RL part chooses best possible weights for AHP to optimize slice specific KPIs, depending on the prevailing wireless environment. Simulation results show that our proposed algorithm outperforms three widely accepted related work in terms of the major KPIs for all three slices. Moreover, our results show that the RL+AHP outperforms the existing DRL based approaches in terms of CPU usage when the number of criteria is reasonably low ($<6$). 

\bibliographystyle{IEEEtran}
\bibliography{ref.bib}
\end{document}